\documentclass[ALICE,manyauthors]{cernphprep}
\usepackage[comma,square,numbers,sort&compress]{natbib}
\usepackage{hyperref}
\usepackage{lineno}
\usepackage{xspace}

\usepackage[T1]{fontenc} 
\usepackage{setspace}
\usepackage{capt-of}
\usepackage{epstopdf}
\usepackage{color}
\usepackage{multirow}
\usepackage{tabu}
 \usepackage[draft]{todonotes}
 \usepackage{float}

\usepackage{tabularx}
\usepackage{changepage}     
\usepackage[figuresright]{rotating}
\usepackage{graphicx} 

\newcolumntype{?}{!{\vrule width 2pt}}
\newcolumntype{@}{!{\vrule width 1.5pt}}

\begin{document}
%

\newcommand{\pp}           {pp\xspace}
\newcommand{\ppbar}        {\mbox{$\mathrm {p\overline{p}}$}\xspace}
\newcommand{\XeXe}         {\mbox{Xe--Xe}\xspace}
\newcommand{\PbPb}         {\mbox{Pb--Pb}\xspace}
\newcommand{\pA}           {\mbox{pA}\xspace}
\newcommand{\pPb}          {\mbox{p--Pb}\xspace}
\newcommand{\AuAu}         {\mbox{Au--Au}\xspace}
\newcommand{\dAu}          {\mbox{d--Au}\xspace}

\newcommand{\s}            {\ensuremath{\sqrt{s}}\xspace}
\newcommand{\snn}          {\ensuremath{\sqrt{s_{\mathrm{NN}}}}\xspace}
\newcommand{\pt}           {\ensuremath{p_{\rm T}}\xspace}
\newcommand{\meanpt}       {$\langle p_{\mathrm{T}}\rangle$\xspace}
\newcommand{\ycms}         {\ensuremath{y_{\rm CMS}}\xspace}
\newcommand{\ylab}         {\ensuremath{y_{\rm lab}}\xspace}
\newcommand{\etarange}[1]  {\mbox{$\left | \eta \right |~<~#1$}}
\newcommand{\yrange}[1]    {\mbox{$\left | y \right |~<~#1$}}
\newcommand{\dndy}         {\ensuremath{\mathrm{d}N_\mathrm{ch}/\mathrm{d}y}\xspace}
\newcommand{\dndeta}       {\ensuremath{\mathrm{d}N_\mathrm{ch}/\mathrm{d}\eta}\xspace}
\newcommand{\avdndeta}     {\ensuremath{\langle\dndeta\rangle}\xspace}
\newcommand{\dNdy}         {\ensuremath{\mathrm{d}N_\mathrm{ch}/\mathrm{d}y}\xspace}
\newcommand{\Npart}        {\ensuremath{N_\mathrm{part}}\xspace}
\newcommand{\Ncoll}        {\ensuremath{N_\mathrm{coll}}\xspace}
\newcommand{\dEdx}         {\ensuremath{\textrm{d}E/\textrm{d}x}\xspace}
\newcommand{\RpPb}         {\ensuremath{R_{\rm pPb}}\xspace}

\newcommand{\nineH}        {$\sqrt{s}~=~0.9$~Te\kern-.1emV\xspace}
\newcommand{\seven}        {$\sqrt{s}~=~7$~Te\kern-.1emV\xspace}
\newcommand{\twoH}         {$\sqrt{s}~=~0.2$~Te\kern-.1emV\xspace}
\newcommand{\twosevensix}  {$\sqrt{s}~=~2.76$~Te\kern-.1emV\xspace}
\newcommand{\five}         {$\sqrt{s}~=~5.02$~Te\kern-.1emV\xspace}
\newcommand{\twosevensixnn}{$\sqrt{s_{\mathrm{NN}}}~=~2.76$~Te\kern-.1emV\xspace}
\newcommand{\fivenn}       {$\sqrt{s_{\mathrm{NN}}}~=~5.02$~Te\kern-.1emV\xspace}
\newcommand{\LT}           {L{\'e}vy-Tsallis\xspace}
\newcommand{\GeVc}         {Ge\kern-.1emV/$c$\xspace}
\newcommand{\MeVc}         {Me\kern-.1emV/$c$\xspace}
\newcommand{\TeV}          {Te\kern-.1emV\xspace}
\newcommand{\MeV}          {Me\kern-.1emV\xspace}
\newcommand{\GeVmass}      {Ge\kern-.2emV/$c^2$\xspace}
\newcommand{\MeVmass}      {Me\kern-.2emV/$c^2$\xspace}
\newcommand{\lumi}         {\ensuremath{\mathcal{L}}\xspace}

\newcommand{\ITS}          {\rm{ITS}\xspace}
\newcommand{\TOF}          {\rm{TOF}\xspace}
\newcommand{\ZDC}          {\rm{ZDC}\xspace}
\newcommand{\ZDCs}         {\rm{ZDCs}\xspace}
\newcommand{\ZNA}          {\rm{ZNA}\xspace}
\newcommand{\ZNC}          {\rm{ZNC}\xspace}
\newcommand{\SPD}          {\rm{SPD}\xspace}
\newcommand{\SDD}          {\rm{SDD}\xspace}
\newcommand{\SSD}          {\rm{SSD}\xspace}
\newcommand{\TPC}          {\rm{TPC}\xspace}
\newcommand{\TRD}          {\rm{TRD}\xspace}
\newcommand{\VZERO}        {\rm{V0}\xspace}
\newcommand{\VZEROA}       {\rm{V0A}\xspace}
\newcommand{\VZEROC}       {\rm{V0C}\xspace}
\newcommand{\Vdecay} 	   {\ensuremath{V^{0}}\xspace}

\newcommand{\ee}           {\ensuremath{e^{+}e^{-}}} 
\newcommand{\pip}          {\ensuremath{\pi^{+}}\xspace}
\newcommand{\pim}          {\ensuremath{\pi^{-}}\xspace}
\newcommand{\kap}          {\ensuremath{\rm{K}^{+}}\xspace}
\newcommand{\kam}          {\ensuremath{\rm{K}^{-}}\xspace}
\newcommand{\pbar}         {\ensuremath{\rm\overline{p}}\xspace}
\newcommand{\kzero}        {\ensuremath{{\rm K}^{0}_{\rm{S}}}\xspace}
\newcommand{\lmb}          {\ensuremath{\Lambda}\xspace}
\newcommand{\almb}         {\ensuremath{\overline{\Lambda}}\xspace}
\newcommand{\Om}           {\ensuremath{\Omega^-}\xspace}
\newcommand{\Mo}           {\ensuremath{\overline{\Omega}^+}\xspace}
\newcommand{\X}            {\ensuremath{\Xi^-}\xspace}
\newcommand{\Ix}           {\ensuremath{\overline{\Xi}^+}\xspace}
\newcommand{\Xis}          {\ensuremath{\Xi^{\pm}}\xspace}
\newcommand{\Oms}          {\ensuremath{\Omega^{\pm}}\xspace}
\newcommand{\degree}       {\ensuremath{^{\rm o}}\xspace}
\newcommand{\trento}{T\raisebox{-.5ex}{R}ENTo}

\newcommand{\vo}{$\rm{V^{0}}$}

\newcommand{\pT}{$p_{\mathrm{T}}$}

\newcommand{\pTnq}{$p_{\mathrm{T}}/n_{q}$}
\newcommand{\pTmT}{$(m_{\mathrm{T}}-m_{0})/n_{q}$}

\newcommand{\vtwo}{$v_{2}$}
\newcommand{\vthree}{$v_{3}$}
\newcommand{\vfour}{$v_{4}$}
\newcommand{\vfive}{$v_{5}$}
\newcommand{\vn}{$v_{n}$}

\newcommand{\etagap}{$|\Delta\eta|>0$}

\newcommand{\pion}{$\pi^{\pm}$}
\newcommand{\kaon}{$\rm{K}^{\pm}$}
\newcommand{\proton}{$\rm{p+\overline{p}}$}
\newcommand{\Ks}{$\rm{K^{0}_{S}}$}
\newcommand{\lambdas}{$\Lambda+\overline{\Lambda}$}

\newcommand{\GeV}{${\rm GeV}/c$}
\newcommand{\sNN}{$\sqrt{s_{\mathrm{NN}}}=5.02$ TeV}

\newcommand{\red}[1]{{\color{red}{#1}}}

\newcommand{\minv}{${\it M}_{\rm{inv}}$}

\begin{titlepage}
\PHyear{2019}       
\PHnumber{271}      
\PHdate{02 December}  

\title{Non-linear flow modes of identified particles in Pb--Pb collisions at \sNN}
\ShortTitle{Non-linear flow modes of identified particles in Pb--Pb collisions}   

\Collaboration{ALICE Collaboration\thanks{See Appendix~\ref{app:collab} for the list of collaboration members}}
\ShortAuthor{ALICE Collaboration} 

\begin{abstract}
\noindent The $p_{\mathrm{T}}$-differential non-linear flow modes, $v_{4,22}$, $v_{5,32}$, $v_{6,33}$ and $v_{6,222}$ for \pion, \kaon, \Ks, \proton, \lambdas~and $\phi$-meson have been measured for the first time at \sNN~in Pb--Pb collisions with the ALICE detector at the Large Hadron Collider. The results were obtained with a multi-particle technique, correlating the identified hadrons with reference charged particles from a different pseudorapidity region. 
These non-linear observables probe the contribution from the second and third order initial spatial anisotropy coefficients to higher flow harmonics. All the characteristic features observed in previous $p_{\mathrm{T}}$-differential anisotropic flow measurements for various particle species are also present in the non-linear flow modes, i.e. increase of magnitude with increasing centrality percentile, mass ordering at low $p_{\mathrm{T}}$ and particle type grouping in the intermediate $p_{\mathrm{T}}$ range. Hydrodynamical calculations (iEBE-VISHNU) that use different initial conditions and values of shear and bulk viscosity to entropy density ratios are confronted with the data at low transverse momenta. These calculations exhibit a better agreement with the anisotropic flow coefficients than the non-linear flow modes. These observations indicate that non-linear flow modes can provide additional discriminatory power in the study of initial conditions as well as new stringent constraints to hydrodynamical calculations.

\end{abstract}
\end{titlepage}

\setcounter{page}{2} 
\tableofcontents
\newpage
\setcounter{page}{3}
difie\section{Introduction}
\label{Sec:Introduction}

Lattice quantum chromodynamics (QCD) calculations \cite{Borsanyi:2010cj,Bhattacharya:2014ara} suggest that at extremely high temperature and energy density a state of matter is produced in which quarks and gluons are no longer confined into hadrons. This state of matter is called the quark-gluon plasma (QGP) \cite{Shuryak:1984nq, Cleymans:1985wb, Bass:1998vz}. The main goal of heavy-ion collision experiments is to study the properties of the QGP, such as the speed of sound, the equation of state and its shear and bulk viscosities.

One of the observables sensitive to these properties is the azimuthal angular distribution of particles emitted in the plane perpendicular to the beam axis. In a heavy-ion collision, the overlap region of the colliding nuclei exhibits an irregular shape \cite{Miller:2003kd,Bhalerao:2006tp, Alver:2008zza, Alver:2010gr, Alver:2010dn, Manly:2005zy, Voloshin:2006gz}. This spatial irregularity is a superposition of the geometry, i.e. centrality \cite{Adam:2015ptt} of the collision reflected in the value of the impact parameter, and the initial energy density in the transverse plane which fluctuates from event to event. Through interactions between partons and at later stages between the produced particles, this spatial irregularity is transferred into an anisotropy in momentum space. The latter is usually decomposed into a Fourier expansion of the azimuthal particle distribution \cite{Voloshin:1994mz} according to

\begin{equation}
\frac{{\rm d}N}{{\rm d}\varphi} \propto 1+2\sum_{{\rm n}=1}^{\infty} v_{\rm n}(p_{\mathrm{T}},\eta) \cos[{\rm n}(\varphi - \Psi_{\rm n})],
\label{Eq:Fourier}
\end{equation}

\noindent where $N$, $p_{\mathrm{T}}$, $\eta$ and $\varphi$ are the particle yield, transverse momentum, pseudorapidity and azimuthal angle of particles, respectively, and $\Psi_{\rm n}$ is the azimuthal angle of the ${\rm n}^{\mathrm{th}}$-order symmetry plane~\cite{Voloshin:2006gz,Bhalerao:2006tp,Alver:2008zza,Alver:2010gr,Alver:2010dn}. The coefficient $v_{\rm n}$ is the magnitude of the ${\rm n}^{\mathrm{th}}$-order flow vector coefficient $V_{\rm n}$, defined as $V_{\rm n} = v_{\rm n}e^{i{\rm n}\Psi_{\rm n}}$, and can be calculated according to 

\begin{equation}
v_{\rm n} = \langle{\cos[{\rm n}(\varphi - \Psi_{\rm n})]}\rangle,
\label{Eq:vn}
\end{equation}

where the angle brackets denote an average over all particles in all events. Since the symmetry planes are not accessible experimentally, the flow coefficients are estimated solely from the azimuthal angles of the particles emitted in the transverse plane. Measurements of different anisotropic flow coefficients at both the Relativistic Heavy Ion Collider (RHIC) \cite{Adams:2003am,Abelev:2007qg,Adler:2003kt,Adare:2006ti,Alver:2007qw,Adcox:2002ms,Adamczyk:2013gw,Adler:2004cj,Afanasiev:2009wq,Adare:2011tg,Ackermann:2000tr,Adler:2001nb,Adler:2002ct,Adler:2002pu,Adams:2003zg,Adams:2004wz,Adams:2004bi} and the Large Hadron Collider (LHC) \cite{Aamodt:2010pa, ALICE:2011ab, Abelev:2012di, Chatrchyan:2012xq, Chatrchyan:2012ta, ATLAS:2011ah, ATLAS:2012at,Abelev:2014pua,Adam:2016nfo,Adam:2016izf, Acharya:2018zuq,
Chatrchyan:2013kba, Adam:2015eta, Acharya:2018lmh,Acharya:2018ihu} not only confirmed the production of a strongly coupled quark-gluon plasma (sQGP) but also contributed in constraining the value of the ratio between shear viscosity and entropy density ($\eta/s$) which is very close to the lower limit of $1/4\pi$ conjectured by AdS/CFT \cite{Kovtun:2004de}. In addition, the comparison between experimental data \cite{Adam:2016izf} and viscous hydrodynamical calculations \cite{Niemi:2015voa} showed that higher order flow coefficients and more importantly their transverse momentum dependence are more sensitive probes than lower order coefficients, i.e. \vtwo~and \vthree, to the initial spatial irregularity and its fluctuations \cite{Alver:2010dn}.\\  
This initial state spatial irregularity is usually quantified with the standard (moment-defined) anisotropy coefficients, $\epsilon_{\rm n}$. In the Monte Carlo Glauber model, $\epsilon_{\rm n}$ and its corresponding initial symmetry plane, $\Phi_{\rm n}$ can be calculated from the transverse positions of the nucleons participating in a collision according to \cite{Teaney:2010vd, Alver:2010gr}

\begin{equation}
\epsilon_{\rm n}e^{i{\rm n}\Phi_{\rm n}} = \frac{\langle{r^{\rm n}e^{i{\rm n}\varphi}}\rangle}{\langle r^{\rm n}\rangle}  ({\rm for}~{\rm n}>1),
\label{Eq:epsilonn}
\end{equation}

where the brackets denote an average over the transverse position of all participating nucleons that have an azimuthal angle $\varphi$ and a polar distance from the centre $r$. Model calculations show that $v_2$ and to a large extent, $v_3$ are for a wide range of impact parameters linearly proportional to their corresponding initial spatial anisotropy coefficients, $\epsilon_{2}$ and $\epsilon_{3}$, respectively \cite{Alver:2010gr}, while for larger values of $\rm n$, $v_{\rm n}$ scales with $\epsilon_{n}'$, a cumulant-based definition of initial anisotropic coefficients.  
As an example, the fourth order spatial anisotropy is given by \cite{Teaney:2013dta,Qian:2017ier}
 
\begin{equation}
\epsilon_{4}'e^{i4\Phi'_4} = \epsilon_{4}e^{i4\Phi_4}  + \frac{3\langle{r^{2}}\rangle^{2}}{\langle r^4\rangle}\epsilon_{2}^{2}e^{i4\Phi_2},
\label{Eq:epsilonnprime}
\end{equation}

where the second term in the right hand side of Eq. \ref{Eq:epsilonnprime} reveals a non-linear dependence of $\epsilon_{4}'$ on the lower order $\epsilon_{2}$. This further supports the earlier ideas that the higher order flow vector coefficients, $V_{\rm n}~({\rm n} > 3)$ obtain contributions not only from the linear response of the system to $\epsilon_{\rm n}$, but also a non-linear response proportional to the product of lower order initial spatial anisotropies \cite{Bhalerao:2014xra,Yan:2015jma}. 

In particular, for a single event, $V_{\rm n}$ with $n=4,5,6$ can be decomposed to the linear ($V_{\rm n}^{\mathrm{L}} $) and non-linear ($ V_{\rm n}^{\mathrm{NL}}$) modes according to

\vspace{-0.55cm}
\begin{align}
V_{4} &= V_{4}^{\mathrm{L}} + V_{4}^{\mathrm{NL}} = V_{4}^{\mathrm{L}} + \chi_{4,22}(V_{2})^2, \nonumber \\
V_{5} &= V_{5}^{\mathrm{L}} + V_{5}^{\mathrm{NL}} = V_{5}^{\mathrm{L}} + \chi_{5,32}V_{3}V_{2}, \nonumber \\
V_{6} &= V_{6}^{\mathrm{L}} + V_{6}^{\mathrm{NL}} = V_{6}^{\mathrm{L}} + \chi_{6,222}(V_{2})^3 + \chi_{6,33}(V_{3})^2 + \chi_{6,42}V_{2}V_{4}^{\mathrm{L}},
\label{Eq:V4V5V6}
\end{align}
\vspace{-0.55cm}

where $\chi_{\rm n,mk}$, known as non-linear flow mode coefficients, quantify the contributions of the non-linear modes to the total $V_{\rm n}$ \cite{Yan:2015jma, Acharya:2017zfg}. For simplicity, the magnitude of the total $V_{\rm n}$ will be referred to as anisotropic flow coefficient ($v_{\rm n}$) in the rest of this article. 
The magnitude of the $p_{\rm{T}}$-differential non-linear modes for higher order flow coefficients, $v_{\rm n}^{\rm{NL}}$, can be written as: 

\begin{align}
v_{4,22}(p_{\rm{T}})&= \frac{\langle v_{4}(p_{\rm{T}})v_{2}^{2}\cos(4\Psi_{4}-4\Psi_{2})\rangle}{\sqrt{\langle v_{2}^{4}\rangle}} \approx \langle v_{4}(p_{\rm{T}})\cos(4\Psi_{4}-4\Psi_{2})\rangle, \label{Eq:V422}\\
v_{5,32}(p_{\rm{T}})&= \frac{\langle v_{5}(p_{\rm{T}})v_{3}v_{2}\cos(5\Psi_{5}-3\Psi_{3}-2\Psi_{2})\rangle}{\sqrt{\langle v_{3}^{2}v_{2}^{2}\rangle}} \approx \langle v_{5}(p_{\rm{T}})\cos(5\Psi_{5}-3\Psi_{3}-2\Psi_{2})\rangle, \label{Eq:V532}\\
v_{6,33}(p_{\rm{T}}) &= \frac{\langle v_{6}(p_{\rm{T}})v_{3}^{2}\cos(6\Psi_{6}-6\Psi_{3})\rangle}{\sqrt{\langle v_{3}^{4}\rangle}} \approx \langle v_{6}(p_{\rm{T}})\cos(6\Psi_{6}-6\Psi_{3})\rangle , \label{Eq:V633}\\
v_{6,222}(p_{\rm{T}}) &= \frac{\langle v_{6}(p_{\rm{T}})v_{2}^{3}\cos(6\Psi_{6}-6\Psi_{2})\rangle}{\sqrt{\langle v_{2}^{6}\rangle}} \approx \langle v_{6}(p_{\rm{T}})\cos(6\Psi_{6}-6\Psi_{2})\rangle,
\label{Eq:V6222}
\end{align}

\noindent where brackets denote an average over all events. The approximation is valid assuming a weak correlation between the lower (${\rm n}=2,3$) and higher (${\rm n}>3$) order flow coefficients \cite{Acharya:2017gsw, Bhalerao:2014xra}.

Various measurements of the \pT-differential anisotropic flow, $v_{\rm n}$(\pT), of charged particles \cite{Voloshin:2008dg, ALICE:2011ab, ATLAS:2012at, Chatrchyan:2013kba, Acharya:2018lmh,Acharya:2018ihu} provided a testing ground for model calculations that attempt to describe the dynamical evolution of the system created in heavy-ion collisions. Early predictions showed that the \pT-differential anisotropic flow for different particle species can reveal more information about the equation of state, the role of the highly dissipative hadronic rescattering phase as well as probing particle production mechanisms \cite{Voloshin:1996nv,Huovinen:2001cy}. In order to test these predictions, $v_{\rm n}$(\pT) coefficients were measured for different particle species at RHIC \cite{Adams:2003am,Abelev:2007qg,Adler:2003kt,Adare:2006ti} and at the LHC \cite{Abelev:2014pua,Adam:2015eta,Adam:2016nfo,Acharya:2018zuq}. These measurements reveal a characteristic mass dependence of $v_{\rm n}$(\pT) in the low transverse momentum region (\pT$<3$~\GeVc), a result of an interplay between radial and anisotropic flow, and mass dependent thermal velocities \cite{Voloshin:1996nv,Huovinen:2001cy}.  
In the intermediate \pT~region (3 $\lesssim$ \pT $\lesssim$ 8 \GeVc) the measurements indicate a particle type grouping where baryons have a larger $v_{\rm n}$ than the one of mesons. This feature was explained in a dynamical model where flow develops at the partonic level followed by quark coalescence into hadrons \cite{Voloshin:2002wa,Molnar:2003ff}. In this picture the invariant spectrum of produced particles is proportional to the product of the spectra of their constituents and, in turn, the flow coefficient of produced particles is the sum of the $v_{\rm n}$ values of their constituents. This leads to the so-called number of constituent quarks (NCQ) scaling, observed to hold at an approximate level of $\pm20$\% for $p_{\rm{T}} > 3$ \GeV~\cite{Adare:2006ti,Adare:2012vq,Abelev:2014pua,Adam:2016nfo}.

The measurements of non-linear flow modes in different collision centralities could pose a challenge to hydrodynamic models and have the potential to further constrain both the initial conditions of the collision system and its transport properties, i.e.\,$\eta/s$ and $\zeta/s$ (the ratio between bulk viscosity and entropy density) \cite{Zhu:2016puf, Acharya:2017zfg}. The \pT-dependent non-linear flow modes of identified particles, in particular, allow the effect of late-stage interactions in the hadronic rescattering phase, as well as the effect of particle production to be tested via the coalescence mechanism to the development of the mass ordering at low \pT~and particle type grouping in the intermediate \pT~region, respectively \cite{ALICE:2011ab,Acharya:2018zuq}.

In this article, we report the first results of the $p_{\rm{T}}$-differential non-linear flow modes, i.e. $v_{4,22}$, $v_{5,32}$, $v_{6,33}$ and $v_{6,222}$ for \pion, \kaon, \Ks, \proton, \lambdas~and $\phi$ measured in Pb--Pb collisions at a centre of mass energy per nucleon pair \sNN, recorded by the ALICE experiment \cite{Aamodt:2008zz} at the LHC. The detectors and the selection criteria used in  this analysis are described in Sec. \ref{Sec:ExpSetup} and \ref{Sec:EventTrackIdentification}, respectively. 
The analysis methodology and technique are presented in Sec. \ref{Sec:Analysis method}. In this article, the identified hadron under study and the charged reference particles are obtained from different, non-overlapping pseudorapidity regions. The azimuthal correlations not related to the common symmetry plane (known as non-flow), including the effects arising from jets, resonance decays and quantum statistics correlations, are suppressed by using multi-particle correlations as explained in Sec. \ref{Sec:Analysis method} and the residual effect is taken into account in the systematic uncertainty as described in Sec. \ref{Sec:Systematics}. All coefficients for charged particles were measured separately for particles and anti-particles and were found to be compatible within statistical uncertainties. The measurements reported in Sec. \ref{Sec:Results} are therefore an average of the results for both charges. The results are reported within the pseudorapidity range $|\eta|<0.8$ for different collision centralities between 0--60\% range of Pb--Pb collisions.

\section{Experimental setup}
\label{Sec:ExpSetup}
ALICE~\cite{Aamodt:2008zz,Abelev:2014ffa} is one of the four large experiments at the LHC, particularly designed to cope with the large charged-particle densities present in central Pb--Pb collisions~\cite{Aamodt:2010pb}. By convention, the $z$-axis is parallel to the beam direction, the $x$-axis is horizontal and points towards the centre of the LHC, and the $y$-axis is vertical and points upwards. The apparatus consists of a set of detectors located in the central barrel, positioned inside a solenoidal magnet which generates a maximum of $0.5$~T field parallel to the beam direction, and a set of forward detectors. 

The Inner Tracking System (\ITS)~\cite{Aamodt:2008zz} and the Time Projection Chamber (\TPC)~\cite{Alme:2010ke} are the main tracking detectors of the central barrel. The \ITS~consists of six layers of silicon detectors employing three different technologies. The two innermost layers, positioned at $r = 3.9$~cm and 7.6~cm,  are Silicon Pixel Detectors (\SPD), followed by two layers of Silicon Drift Detectors (\SDD) ($r = 15$~cm and 23.9~cm). Finally, the two outermost layers are double-sided Silicon Strip Detectors (\SSD) at $r = 38$~cm and 43~cm. The \TPC~has a cylindrical shape with an inner radius of about 85 cm, an outer radius of about 250 cm, and a length of 500 cm and it is positioned around the \ITS. It provides full azimuthal coverage in the pseudorapidity range $|\eta| < 0.9$. 

Charged particles were identified using the information from the \TPC~and the \TOF~detectors~\cite{Aamodt:2008zz}. The \TPC~allows for a simultaneous measurement of the momentum of a particle and its specific energy loss ($\langle \mathrm{d}E/\mathrm{d}x \rangle$) in the gas. The detector provides a separation more than two standard deviations ($2\sigma$) for different hadron species at $p_{\mathrm{T}} < 0.7$~GeV/$c$ and the possibility to identify particles on a statistical basis in the relativistic rise region of $\mathrm{d}E/\mathrm{d}x$ (i.e.~$2 < p_{\rm{T}} < 20$~GeV/$c$)~\cite{Abelev:2014ffa}. The $\mathrm{d}E/\mathrm{d}x$ resolution for the 5$\%$ most central Pb--Pb collisions is 6.5$\%$ and improves for more peripheral collisions~\cite{Abelev:2014ffa}. The \TOF~detector is situated at a radial distance of 3.7 m from the beam axis, around the \TPC~and provides a $3\sigma$ separation between $\pi$--K and K--$\rm{p}$ up to $p_{\mathrm{T}} = $ 2.5~GeV/$c$ and $p_{\mathrm{T}} = 4$~GeV/$c$, respectively~\cite{Abelev:2014ffa}. This is done by measuring the flight time of particles from the collision point with a resolution of about $80$~ps. The start time for the \TOF~measurement is provided by the T0 detectors, two arrays of Cherenkov counters positioned at opposite sides of the interaction points covering $4.6 < \eta < 4.9$ (T0A) and $-3.3 < \eta < -3.0$ (T0C). The start time is also determined using a combinatorial algorithm that compares the timestamps of particle hits measured by the TOF to the expected times of the tracks, assuming a common event time $t_{ev}$. Both methods of estimating the start time are fully efficient for the 80\% most central Pb--Pb collisions \cite{Abelev:2014ffa}.

A set of forward detectors, the \VZERO~scintillator arrays~\cite{Abbas:2013taa}, were used in the trigger logic and for the determination of the collision centrality. The \VZERO~consists of two detectors, the \VZEROA~and the \VZEROC, positioned on each side of the interaction point, covering the pseudorapidity intervals of $2.8 < \eta < 5.1$ and $-3.7 < \eta < -1.7$, respectively. 

For more details on the ALICE apparatus and the performance of the detectors, see Refs.~\cite{Aamodt:2008zz,Abelev:2014ffa}.

\section{Event sample, track selection and particle identification}
\label{Sec:EventTrackIdentification}
\subsection{Trigger selection and data sample}
\label{SubSec:Event}
The analysis is performed on minimum bias Pb--Pb collision data at \sNN~collected by the ALICE detector in 2015. These events were triggered by the coincidence between signals from both V0A and V0C detectors. An offline event selection, exploiting the signal arrival time in V0A and V0C, measured with a 1 ns resolution, was used to discriminate beam induced-background (e.g. beam-gas events) from collision events. This led to a reduction of background events in the analysed samples to a negligible fraction (< 0.1\%) \cite{Abelev:2014ffa}. Events with multiple reconstructed vertices were rejected by comparing multiplicity estimates from the V0 detector to those from the tracking detectors at midrapidity, exploiting the difference in readout times between the systems. The fraction of pileup events left after applying these dedicated pileup removal criteria is negligible. All events selected for the analysis had a reconstructed primary vertex position along the beam axis ($z_{vtx}$) within 10 cm from the nominal interaction point. After all the selection criteria, a filtered data sample of approximately 40 million Pb--Pb events in the 0--60\% centrality interval was analysed to produce the results presented in this article.

Events were classified according to fractions of the inelastic hadronic cross section. The 0--5\% interval represents the most central interactions (i.e. smallest impact parameter) and is referred to as most central collisions. On the other hand, the 50--60\% centrality interval corresponds to the most peripheral (i.e. largest impact parameter) collisions in the analysed sample. The centrality of the collision was estimated using the signal amplitude measured in the V0 detectors which is related to the number of particles crossing their sensitive areas. Details about the centrality determination can be found in \cite{Abelev:2013qoq}.

\subsection{Selection of primary \pion, \kaon~and \proton}
\label{SubSec:Track}
In this analysis, tracks are reconstructed using the information from the TPC and the ITS detectors. The tracking algorithm, based on the Kalman filter \cite{Billoir:1983mz,Billoir:1985nq}, starts from a collection of space points (referred to as clusters) inside the TPC and provides the quality of the fit by calculating its $\chi^{2}$ value. Each space point is reconstructed at one of the TPC pad rows \cite{Aamodt:2008zz}, where the deposited ionisation energy is also measured. The specific ionisation energy loss $\langle{\rm{dE}/dx}\rangle$ is estimated using a truncated mean, excluding the 40\% highest-charge clusters associated to the track. The obtained $\langle{\rm{dE}/dx}\rangle$ has a resolution, which we later refer to as $\sigma_{\rm{TPC}}$. The tracks are propagated to the outer layer of the ITS, and the tracking algorithm attempts to identify space points in each of the consecutive layers, reaching the innermost ones (i.e. SPD). The track parameters are then updated using the combined information from both the TPC and the ITS detectors. 

Primary charged pions, kaons and (anti-)protons were required to have at least 70 reconstructed space points out of the maximum of 159 in the TPC. The average distance between space point and the track fit per TPC space point per degree of freedom (see \cite{Abelev:2014ffa} for details) was required to be below 2. These selections reduce the contribution from short tracks, which are unlikely to originate from the primary vertex. To reduce the contamination by secondary tracks from weak decays or from the interaction with the material, only particles within a maximum distance of closest approach (DCA) between the tracks and the primary vertex in both the transverse plane (${\rm{DCA}}_{xy} < 0.0105 + 0.0350(p_{\rm{T}}~c/{\rm{GeV}})^{-1.1}$ cm) and the longitudinal direction ($\rm{DCA}_{z} < 2$ cm) were analysed. Moreover, the tracks were required to have at least two associated ITS clusters in addition to having a hit in either of the two SPD layers. This selection leads to an efficiency of about 80\% for primary tracks at \pT~$\sim0.6$ \GeV~and a contamination from secondaries of about 5\% at \pT~$=1$ \GeV~\cite{Abelev:2013vea}. These values depend on particle species and transverse momentum \cite{Abelev:2013vea}. 

The particle identification (PID) for pions (\pion), kaons (\kaon) and protons (\proton) used in this analysis relies on the two-dimensional correlation between the number of standard deviations in units of the resolution from the expected signals of the TPC and the TOF detectors similar to what was reported in \cite{Abelev:2014pua,Adam:2016nfo,Acharya:2018zuq}. In this approach particles were selected by requiring their standard deviations from the $\langle d{\it E}/d{\it x}\rangle$ and $t_{\rm TOF}$ values to be less than a $p_{\rm T}$-dependent value, maintaining a minimum purity of 90\% for \pion~and 75\% for \kaon~and 80\% for \proton. In order to further reduce the contamination from other species, the standard deviation of a given track was required to be the minimum among other candidate species. 

In addition, for the evaluation of systematic effects (see Section \ref{Sec:Systematics}) the minimum purity was varied to more strict values, a condition that becomes essential with increasing transverse momentum where the relevant detector response for different particle species starts to overlap. The results for all three particle species were extrapolated to 100\% purity and the uncertainty from the extrapolation was also considered in the estimation of the total systematic uncertainty.

\subsection{Reconstruction of \Ks, \lambdas~and $\phi$ meson}
\label{SubSec:K0sLambdaPhiRec}

In this analysis, the \Ks~and \lambdas~are reconstructed via the following fully hadronic decay channels: \Ks $\rightarrow \pi^{+} + \pi^{-}$ and  $\Lambda(\overline{\Lambda})\rightarrow {\rm p}(\overline{\rm p})+\pi^{-}(\pi^{+})$ with branching ratios of 69.2\% and 63.9\% \cite{PhysRevD.98.030001}, respectively. The reconstruction is performed by identifying the candidates of secondary vertices, denoted as \vo s, from which two oppositely-charged decay products originate. Such candidates are obtained during data processing by looking for a characteristic V-shaped decay topology among pairs of reconstructed tracks.

The daughter tracks were reconstructed within $|\eta|<0.8$, while the criteria on the number of TPC space points, the number of crossed TPC pad rows, and the percentage of the expected TPC space points used to reconstruct a track are identical to those applied for primary particles. In addition, the minimum DCA of the daughter tracks to the primary vertex is 0.1 cm. Furthermore, the maximum DCA of the daughter tracks is 0.5 cm to ensure that they are products of the same decay. To suppress the combinatorial background, the PID is applied for the daughter particles in the whole \pT~region by requiring the particle to be within 3$\sigma_{\rm TPC}$ for a given species hypothesis.

To reject combinatorial background, the cosine of the pointing angle, $\theta_{p}$, was required to be larger than 0.998. This angle is defined as the angle between the momentum vector of the \vo~candidate assessed at its decay vertex and the line connecting the \vo~decay vertex to the primary vertex and has to be close to 1 as a result of momentum conservation. In addition, only the candidates reconstructed between 5 and 100 cm from the nominal primary vertex in radial direction were accepted. The lower value was chosen to avoid any bias from the efficiency loss when secondary tracks are being wrongly matched to clusters in the first layer of the ITS, where the occupancy is the largest. To assess the systematic uncertainty related to the contamination from \lambdas~and electron-positron pairs coming from $\gamma$-conversions to the \Ks~sample, a selection in the Armenteros-Podolanski variables \cite{doi:10.1080/14786440108520416} was applied for the \Ks~candidates, rejecting the ones with $q\le 0.2|\alpha|$. Here $q$ is the momentum projection of the positively charged daughter track in the plane perpendicular to the \vo~momentum and $\alpha = (p_{\rm{L}}^{+} - p_{\rm{L}}^{-})/(p_{\rm{L}}^{+} + p_{\rm{L}}^{-})$ with $p_{\rm{L}}^{\pm}$ the projection of the positive or negative daughter track momentum onto the momentum of the \vo. 

The reconstruction of $\phi$ meson candidates is done via the hadronic decay channel: $\phi \rightarrow {\rm K}^{+} + {\rm K}^{-}$ with a branching ratio of 48.9\% \cite{PhysRevD.98.030001}. The $\phi$ meson candidates were reconstructed from the charged tracks passing all criteria for charged kaons. These kaon daughters were identified utilising the Bayesian PID approach \cite{Adam:2016acv} with a minimum probability threshold of $85\%$ using the TPC and TOF detectors. Additionally, to reduce combinatorial background, a track was identified as a kaon if it had the highest probability among all considered species ($e^{\pm}$,~$\mu^{\pm}$,~\pion,~\kaon~and \proton). The vector sum of all possible pairs of charged kaons are called $\phi$~candidates. The invariant mass distribution (${\it M}_{\rm inv}^{\rm K^{+}K^{-}}$) of $\phi$~candidates was then obtained in various \pT~intervals by subtracting a combinatorial background yield from the candidate yield. This combinatorial background yield was estimated from like-sign kaon pairs (unphysical $\phi$ state with total charge of $\pm2$) normalised to the candidate yield. 

\section{Analysis method}
\label{Sec:Analysis method}
In this article the \pT-differential non-linear flow modes are calculated based on Eqs. \ref{Eq:V422}-\ref{Eq:V6222}. Each event is divided into two subevents ``$\rm{A}$'' and ``$\rm{B}$'', covering the ranges $-0.8< \eta < 0.0$ and $0.0 <\eta< 0.8$, respectively. Thus $v_{\rm n,mk}(p_{\rm{T}})$ is a weighted average of $v_{\rm n,mk}^{\rm{A}}(p_{\rm{T}})$ and $v_{\rm n,mk}^{\rm{B}}(p_{\rm{T}})$. The measured $v_{\rm n,mk}^{\rm{A(B)}}(p_{\rm{T}})$ coefficients are calculated using $d_{\rm n,mk}(p_{\rm{T}})$ and $c_{\rm mk,mk}$ multi-particle correlators given by

\begin{equation}
d_{\rm n,mk}(p_{\rm{T}}) = \langle v_{\rm n}(p_{\rm{T}})v_{\rm m}v_{\rm k}\cos(n\Psi_{\rm n}-m\Psi_{\rm m}-k\Psi_{\rm k}) \rangle,
\label{Eq:dnmki}
\end{equation}

\begin{equation}
c_{\rm mk,mk} = \langle v_{\rm m}^{2}v_{\rm k}^{2}\rangle.
\label{Eq:cmkimki}
\end{equation}

These correlators were obtained using the Generic Framework with sub-event method originally used in \cite{Bilandzic:2013kga,Acharya:2017zfg,Pacik:2711398}, which allows precise non-uniform acceptance and efficiency corrections. In this analysis, $d_{\rm n,mk}(p_{\rm{T}})$ is measured by correlating the azimuthal angle of the particle of interest ($\varphi_{1}(p_{\rm T})$) from subevent ``$\rm{A}$''(``$\rm{B}$'') with that of reference particles\footnote{Later in the text particle of interest and reference particles will be referred to as POI and RFP, respectively.} from subevent ``$\rm{B}$''(``$\rm{A}$'') and $c_{\rm mk,mk}$ by selecting half of the reference particles from subevent ``$\rm{A}$'' and the other half from ``$\rm{B}$''. Thus, Eqs.\ref{Eq:V422} to \ref{Eq:V6222} for $v_{\rm n,mk}^{\rm{A}}(p_{\rm{T}})$ translate to

\begin{align}
v_{4,22}^{\rm{A}}(p_{\rm{T}}) &= \frac{d_{4,22}^{\rm{A}}(p_{\rm{T}})}{\sqrt{c_{22,22}}} =  \frac{\langle\langle \cos(4\varphi^{\rm{A}}_{1}(p_{\rm{T}})-2\varphi^{\rm{B}}_{2}-2\varphi^{\rm{B}}_{3})\rangle\rangle}{\sqrt{\langle\langle \cos(2\varphi^{\rm{A}}_{1}+2\varphi^{\rm{A}}_{2}-2\varphi^{\rm{B}}_{3}-2\varphi^{\rm{B}}_{4}) \rangle\rangle}}, \label{Eq:VA422} \\
v_{5,32}^{\rm{A}}(p_{\rm{T}}) &= \frac{d_{5,32}^{\rm{A}}(p_{\rm{T}})}{\sqrt{c_{32,32}}} = \frac{\langle\langle \cos(5\varphi^{\rm{A}}_{1}(p_{\rm{T}})-3\varphi^{\rm{B}}_{3}-2\varphi^{\rm{B}}_{2})\rangle\rangle}{\sqrt{\langle\langle \cos(3\varphi^{\rm{A}}_{1}+2\varphi^{\rm{A}}_{2}-3\varphi^{\rm{B}}_{3}-2\varphi^{\rm{B}}_{4}) \rangle\rangle}}, \label{Eq:VA532}\\
v_{6,33}^{\rm{A}}(p_{\rm{T}}) &= \frac{d_{6,33}^{\rm{A}}(p_{\rm{T}})}{\sqrt{c_{33,33}}} =\frac{\langle\langle \cos(6\varphi^{A}_{1}(p_{\rm{T}})-3\varphi^{\rm{B}}_{2}-3\varphi^{\rm{B}}_{3})\rangle\rangle}{\sqrt{\langle\langle \cos(3\varphi^{\rm{A}}_{1}+3\varphi^{\rm{A}}_{2}-3\varphi^{\rm{B}}_{3}-3\varphi^{\rm{B}}_{4}) \rangle\rangle}}, \label{Eq:VA633}\\
v_{6,222}^{\rm{A}}(p_{\rm{T}}) &= \frac{d_{6,222}^{\rm{A}}(p_{\rm{T}})}{\sqrt{c_{222,222}}} =\frac{\langle\langle \cos(6\varphi^{\rm{A}}_{1}(p_{\rm{T}})-2\varphi^{\rm{B}}_{2}-2\varphi^{\rm{B}}_{3}-2\varphi^{\rm{B}}_{4})\rangle\rangle}{\sqrt{\langle\langle \cos(2\varphi^{\rm{A}}_{1}+2\varphi^{\rm{A}}_{2}+2\varphi^{\rm{A}}_{3}-2\varphi^{\rm{B}}_{4}-2\varphi^{\rm{B}}_{5}-2\varphi^{\rm{B}}_{6}) \rangle\rangle}},
\label{Eq:VA6222}
\end{align}

where $\langle\langle\rangle\rangle$ denotes an average over all particles and events.
This multi-particle correlation technique by construction removes a significant part of non-flow correlations. In order to further reduce residual non-flow contributions, a pseudorapidity gap was applied between the two pseudorapidity regions ($|\Delta\eta|>0.4$). In addition, particles with like-sign charges were correlated. These two variations do not significantly affect the results but any variation was included in the final systematics in Tab. \ref{SystematicsValues:PID}.

For charged hadrons, i.e.\,\pion, \kaon~and \proton, the $d_{\rm n,mk}$ correlators are calculated on a track-by-track basis as a function of \pT~for each centrality percentile. For particle species reconstructed on a statistical basis from the decay products, i.e.\,\Ks, \lambdas~and $\phi$ meson, the selected sample contains both signal and the background. Therefore, the $d_{\rm n,mk}$ correlators are measured as a function of invariant mass (${\it M}_{\rm inv}$)~and \pT~for each centrality percentile. The $d_{\rm n,mk}$ vs. \minv~method is based on the additivity of correlations and is a weighted sum of the $d_{\rm n,mk}^{\rm{sig}}$ and $d_{\rm n,mk}^{\rm{bkg}}$ according to

\begin{equation}
d_{\rm n,mk}^{\rm total}({\it M}_{\rm inv}, p_{\rm T}) = \frac{N^{\rm sig}}{N^{\rm sig}+N^{\rm bkg}}({\it M}_{\rm inv}, p_{\rm T})\,d_{\rm n,mk}^{\rm sig}(p_{\rm T})+\frac{N^{\rm bkg}}{N^{{\rm sig}}+N^{{\rm bkg}}}({\it M}_{\rm inv}, p_{\rm T})\,d_{\rm n,mk}^{\rm bkg}({\it M}_{\rm inv}, p_{\rm T}),
\label{Eq:dnmk}
\end{equation}

where $N^{\rm{sig}}$ and $N^{\rm{bkg}}$ are signal and background yields obtained for each \pT~interval and centrality percentile from fits to the \Ks, \lambdas~and $\phi$ meson invariant mass distributions. To obtain the \pT-differential yield of \Ks~and \lambdas, the invariant mass distributions at various \pT~intervals were parametrised as a sum of two Gaussian distributions and a third-order polynomial function. The latter was introduced to account for residual contamination (background yield) that is present in the \Ks~and \lambdas~signals after the topological and daughter track selections. The \Ks~and \lambdas~yields were extracted by integration of the Gaussian distribution. The obtained yields were not corrected for feed-down from higher mass baryons ($\Xi^{\pm}$,$\Omega^{\pm}$) as earlier studies have shown that these have a negligible effect on the measured $v_{\rm n}$ \cite{Abelev:2014pua}. Similarly, to obtain the \pT-differential yield of $\phi$-mesons, the invariant mass distributions of the candidate yield was parametrized as a sum of a Breit-Wigner distribution and a third-order polynomial function, the latter introduced to account for residual contamination.

\begin{figure}[!htb]
\begin{center}
\includegraphics[scale=0.45]{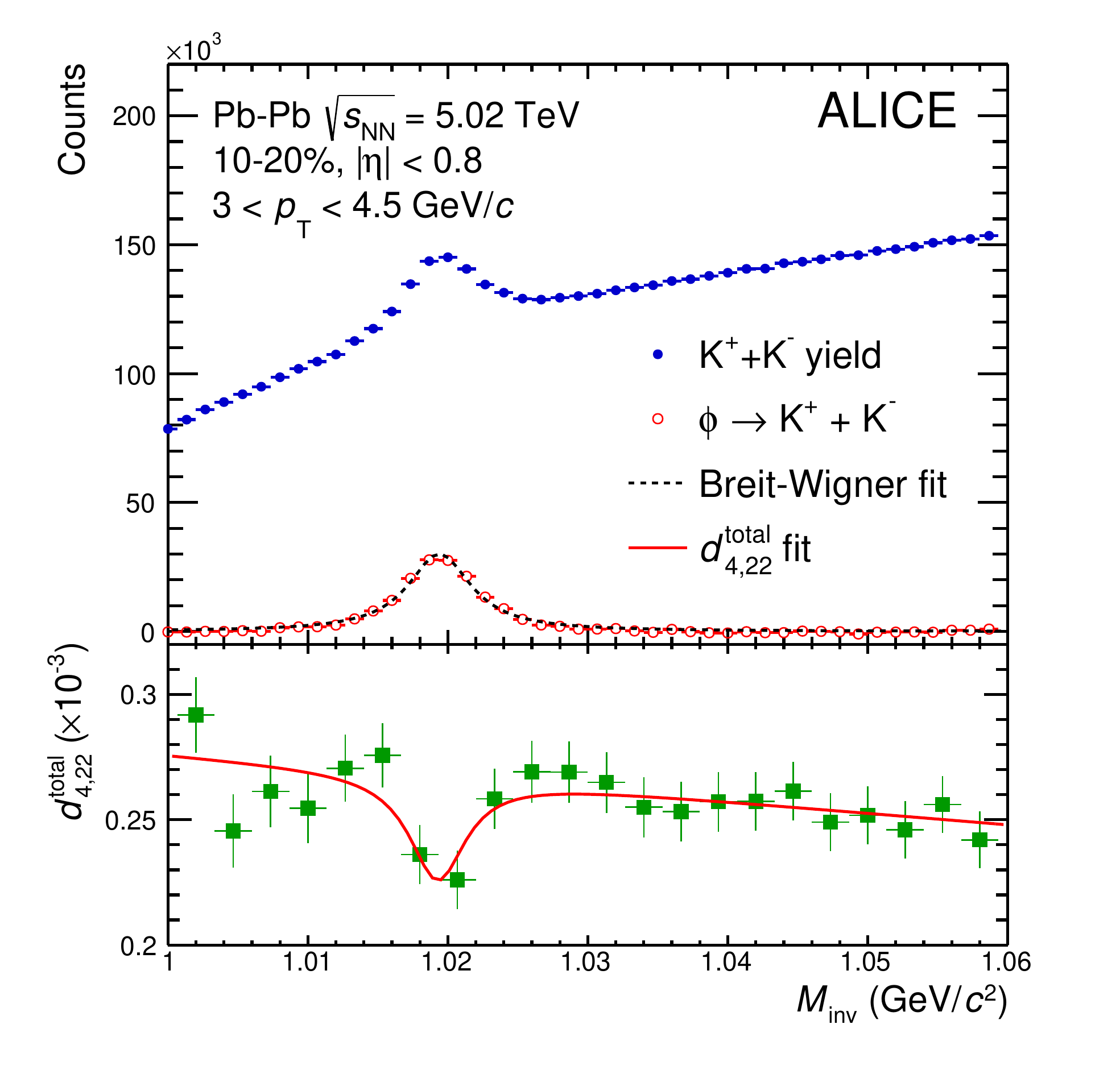}
\end{center}
\caption{Reconstruction and $d_{4,22}$ measurement of $\phi$-mesons. Upper panel: extraction of $N^{\rm{sig}}$ and $N^{\rm{bkg}}$ by fitting the invariant mass (${\it M}_{\rm{inv}}$) distribution for $\phi$-meson candidates from pairs of kaons with opposite charges for $3<p_{\rm{T}}<4.5~{\rm GeV}/c$ and the 10--20\% centrality interval, lower panel: extraction of $d_{4,22}^{\rm sig}$ by fitting Eq. \ref{Eq:dnmk} to the invariant mass dependence of $d_{4,22}^{\rm total}$.}
\label{d422_phi_meson}
\end{figure}

To extract $d_{\rm n,mk}^{\rm{sig}}$ in a given \pT~range, $d_{\rm n,mk}^{\rm total}({\it M}_{\rm inv})$ was fitted together with the fit values from the invariant mass distribution and parametrising $d_{\rm n,mk}^{\rm{bkg}}({\it M}_{\rm inv})$ with a first order polynomial function. Figure \ref{d422_phi_meson} illustrates this procedure for the $\phi$-meson, with the invariant mass distribution in the upper panel and the measurement of $d_{4,22}^{\rm total}({\it M}_{\rm inv})$ in the lower panel. 

\section{Systematic uncertainties}
\label{Sec:Systematics}

The systematic uncertainties were estimated by varying the selection criteria for all particle species as well as the topological reconstruction requirements for \Ks, \lambdas~and $\phi$. The contributions from different sources were extracted from the relative ratio of the \pT-differential $v_{n,mk}$ between the default selection criteria described in Sec.~\ref{Sec:EventTrackIdentification} and their variations summarised in this section. Sources with a statistically significant contribution (where significance is evaluated as recommended in \cite{Barlow:2002yb}) were added in quadrature to form the final value of the systematic uncertainties on the non-linear flow modes. 
An overview of the magnitude of the relative systematic uncertainties per particle species is given in Tab. \ref{SystematicsValues:PID} for \pion, \kaon~and \proton~and Tab. \ref{SystematicsValues:V0} for \Ks, \lambdas~and the $\phi$-meson. The systematic uncertainties are grouped into five categories, i.e.\,event selection, tracking, particle identification, topological cuts and non-flow contribution and are described below.

The effects of event selection criteria on the measurements were studied by:  (i) varying the primary vertex position along the beam axis ($z_{vtx}$) from a nominal $\pm$10 cm to $\pm$8 cm and $\pm$6 cm; (ii) changing the centrality estimator from the signal amplitudes in the V0 scintillator detectors to the number of clusters in the first or second layer of the SPD, (iii) analysing events recorded for different magnetic field polarities independently; (iv) not rejecting all events with tracks caused by pileup. 

Systematic uncertainties induced by the selection criteria imposed at the track level were investigated by: (i) changing the tracking from global mode, where combined track information from both TPC and ITS detectors are used, to what is referred to as hybrid mode. In the latter mode, track parameters from the TPC are used if the algorithm is unable to match the track reconstructed in the TPC with associated ITS clusters; (ii) increasing the number of TPC space points from 60 up to 90 and (iii) decreasing the value of the $\chi^{2}$ per TPC space point per degree of freedom from 4 to 3; (iv) varying the selection criteria on both the transverse and longitudinal components of the DCA to estimate the impact of secondary particles from a strict \pT-dependent cut to 0.15 cm and 2 cm to 0.2 cm, respectively.

Systematic uncertainties associated with the particle identification procedure were studied by varying the PID method from a \pT-dependent one described in Sec. \ref{SubSec:Track} to an even stricter version where the purity increases to higher than 95\% (\pion), 80\% (\kaon) and 80\% (\proton) across the entire \pT~range of study. The second approach relied on the Bayesian method with a probability of at least 80\% which gives an increase in purity to at least 97\% (\pion), 87\% (\kaon) and 90\% (\proton) across the entire \pT~range of study. To further check the effect of contamination the purity of these species was extrapolated to 100\%. 

The topological cuts were also varied to account for the \vo~and $\phi$-meson reconstruction. These selection criteria were varied by 
(i) changing the reconstruction method for  \vo~particles to an alternate technique that uses raw tracking information during the Kalman filtering stage (referred to as online V0 finder); (ii) varying the minimum radial distance from the primary vertex at which the \vo~can be produced from 5 cm to 10 cm; (iii) changing the minimum value of the cosine of pointing angle from 0.998 to 0.99; (iv) varying the minimum number of crossed TPC pad rows by the \vo~daughter tracks from 70 to 90; (v) changing the requirement on the minimum number of TPC space points that are used in the reconstruction of the \vo~daughter tracks form 70 to 90; (vi) requesting a minimum ratio of crossed to findable TPC clusters from 0.8 to 1.0; (vii) changing the minimum DCA of the \vo~daughter tracks to the primary vertex from 0.1 cm to 0.3 cm; (viii) changing the maximum DCA of the \vo~daughter tracks from 0.5 cm to 0.3 cm; (ix) requiring a minimum \pT~of the \vo~daughter tracks of 0.2 \GeV. 

In addition, the non-flow contribution was studied by (i) selecting like sign pairs of particles of interest and reference particles to decrease the effect from the decay of resonance particles; (ii) applying pseudorapidity gaps between the two subevents from $|\Delta\eta|>0.0$ to $|\Delta\eta|>0.4$.

Tables \ref{SystematicsValues:PID} and \ref{SystematicsValues:V0} summarise the maximum relative systematic uncertainties for each individual systematic source described above for all transverse momenta. The systematic uncertainties are expressed for each non-linear mode and particle species in a range to account for all centrality intervals in this article. 

\begin{table}[!ht]
\caption{List of the maximum relative systematic uncertainties of each individual source for $v_{\rm n,mk}$ of \pion, \kaon~and \proton. The uncertainties depend on the transverse momenta. Percentage ranges are given to account for all centrality intervals.}
\resizebox{\textwidth}{!}{\begin{tabular}{ |p{4.5cm} |l|c|c|c|c|c|c|c|c|c|c|c|c|}
\hline
\multicolumn{1}{| c |}{} & \multicolumn{3}{| c |}{ $v_{4,22}$ } & \multicolumn{3}{| c |}{ $v_{5,32}$} & \multicolumn{3}{| c |}{ $v_{6,33}$} & \multicolumn{3}{| c |}{ $v_{6,222}$} \\
\hline
Uncertainty source  & \pion &  \kaon & \proton &  \pion & \kaon & \proton &  \pion &  \kaon & \proton &  \pion &  \kaon & \proton \\ \hline  \hline
Primary $z_{vtx}$  & 0--2\% & 1--3\% & 0--3\% & 0--3\% & 1--3\% & 1--4\% & 3--5\% & 2--5\% & 3--5\% & 2--7\% & 2--7\% & 4--7\%\\
Centrality estimator  & 0--4\% & 1--4\% & 1--5\% & 0--4\% & 1--3\% & 2--4\% & 4--10\% & 4--10\% & 5--10\% & 3--10\% & 5--10\% & 4--10\%\\
Magnetic field polarity & 0--2\% & 0--3\% & 0--3\% & 0--4\% & 0--5\% & 0--5\% & 0--10\% & 0--10\% & 0--10\% & 0--10\% & 0--10\% & 0--10\% \\
Pileup rejection & 0--4\% & 0--3\% & 0--4\% & 0--5\% & 1--5\% & 0--5\% & 5--7\% & 5--10\% & 5--8\% & 4--10\% & 4--10\% & 2--10\%\\\hline
Tracking mode  & 1--4\% & 1--5\% & 1--4\% & 2--6\% & 3--5\% & 2--8\% & 0--8\% & 0--7\% & 3--8\% & 1--10\% & 4--10\% & 2--10\% \\
Number of \TPC~space points &  1--2\% & 0--2\% & 0--2\% & 0--3\% & 1--3\% & 1--3\% & 4--8\% & 3--8\% & 3--8\% & 2--8\% & 4--8\% & 4--8\% \\
$\chi^2$ per \TPC~space point & 0--2\% & 1--2\% & 1--3\% & 1--3\% & 1--3\% & 2--4\% & 3--5\% & 3--6\% & 3--6\% & 2--6\% & 4--7\% & 4--7\%\\
$\rm DCA_{xy}$ & 0--2\% & 0--2\% & 1--3\% & 0--3\% & 1--3\% & 1--3\% & 2--7\% & 2--8\% & 4--8\% & 2--8\% & 4--8\% & 3--8\%\\
$\rm DCA_{z}$ & 0--3\% & 0--2\% & 1--2\% & 1--2\% & 1--3\% & 2--3\% & 3--7\% & 3--7\% & 5--7\% & 2--7\% & 4--8\% & 2--8\%\\\hline
Particle identification & 1--5\% & 1--5\% & 1--3\% & 1--5\% & 2--5\% & 1--5\% & 5--10\% & 5--10\% & 6--12\% & 4--12\% & 6--15\% & 4--15\%\\\hline
POI vs. RFP charges & 0--2\% & 0--3\% & 2--3\% & 0--4\% & 0--4\% & 2--4\% & 0--4\% & 0--6\% & 0--6\% & 0\% & 0\% & 0\% \\ 
$\eta$ gap & 1--3\% & 1--4\% & 1--2\% & 1--4\% & 1--4\% & 1--5\% & 0--5\% & 0--5\% & 0--5\% & 0\% & 0\% & 0\%  \\\hline 
\end{tabular}}
\label{SystematicsValues:PID}
\end{table}

\begin{table}[!ht]
\centering
\caption{List of the maximum relative systematic uncertainties of each individual source for $v_{\rm n,mk}$ of \Ks, \lambdas~and $\phi$-meson. The uncertainties depend on the transverse momenta and centrality interval. Percentage ranges are given to account for all centrality intervals. "N/A" indicates that a certain check was not applicable to the given particle of interest. If a source was checked and proved to have a negligible effect, the field is marked as "--".}
\resizebox{0.75\textwidth}{!}{\begin{tabular}{ |p{6.7cm} |l|c|c|c|c|c|c|c|}
\hline
\multicolumn{1}{| c |}{} & \multicolumn{3}{| c |}{ $v_{4,22}$ } & \multicolumn{2}{| c |}{ $v_{5,32}$} & \multicolumn{2}{| c |}{ $v_{6,33}$} \\
\hline
Uncertainty source  & \Ks &  \lambdas & $\phi$ &  \Ks &  \lambdas & \Ks &  \lambdas   \\ \hline  \hline
Primary $z_{vtx}$  & 0\% & 0-2\% & 1\% & 0\% & 0--3\% & 0\% & 1--3\%\\ \hline
Tracking mode  & - & - & 2\% & - & - & - & - \\
Number of \TPC~space points & 0--3\% & 1--2\% & 2\% & 0\% & 2\% & 0\% & 2\%  \\ \hline
Particle identification & - & - & 4--6\% & - & - & - & -\\\hline
Reconstruction method (\vo~finder) & 3--5\% & 2--3\% & N/A & 5\% & 1\% & 5\% & 1\%   \\
Decay radius & 3--5\% & 1--3\% & N/A & 5--6\% & 0--2\% & 5\% & 2\%\\
Ratio of crossed to findable TPC clusters & 0--2\% & 0--3\% & N/A & 0\%  & 1--2\% & 0\% & 3\%  \\
DCA decay products to primary vertex & 2--5\% & 2--4\% & N/A & 4--5\% & 2--3\% & 5\% & 2--3\%  \\
DCA between decay products & 0--3\% & 1--2\% & N/A & 0--4\% & 0--4\% & 0\% & 0--4\% \\
Pointing angle $\cos(\theta_{\rm p})$ & 3--4\% & 0--2\% & N/A & 3--4\% & 0--3\% & 3\% & 1\%  \\
Minimum \pT~of daughter tracks & 1--3\% & 0--1\% & N/A & 2--3\% & 2--3\% & 0\% & 0--3\%  \\ \hline
\end{tabular}}
\label{SystematicsValues:V0}
\end{table}

\newpage

\section{Results and discussion}
\label{Sec:Results}

In this section, the results of the \pT-dependent non-linear flow modes $v_{4,22}$, $v_{5,32}$, $v_{6,33}$ and $v_{6,222}$ of identified particles are presented for various centrality intervals in Pb--Pb collisions at \sNN. We first present the centrality and \pT~dependence of $v_{\rm n,mk}$ in Sec. \ref{SubSec:pTdependence}. The scaling properties of the non-linear flow modes are also discussed in this section. These results are compared with $v_{\rm n}$ measurements for the same particle species in Sec. \ref{SubSec:comparewithvn}. Finally, the comparison with two model calculations is shown in Sec. \ref{SubSec:hydro}. Note that in some of the following sections the same data are used in different representations to highlight the various physics implications of the measurements in each section.

\subsection{Centrality and \pT~dependence of non-linear flow modes}
\label{SubSec:pTdependence}

\begin{figure}[!htb]
\begin{center}
\includegraphics[scale=0.82]{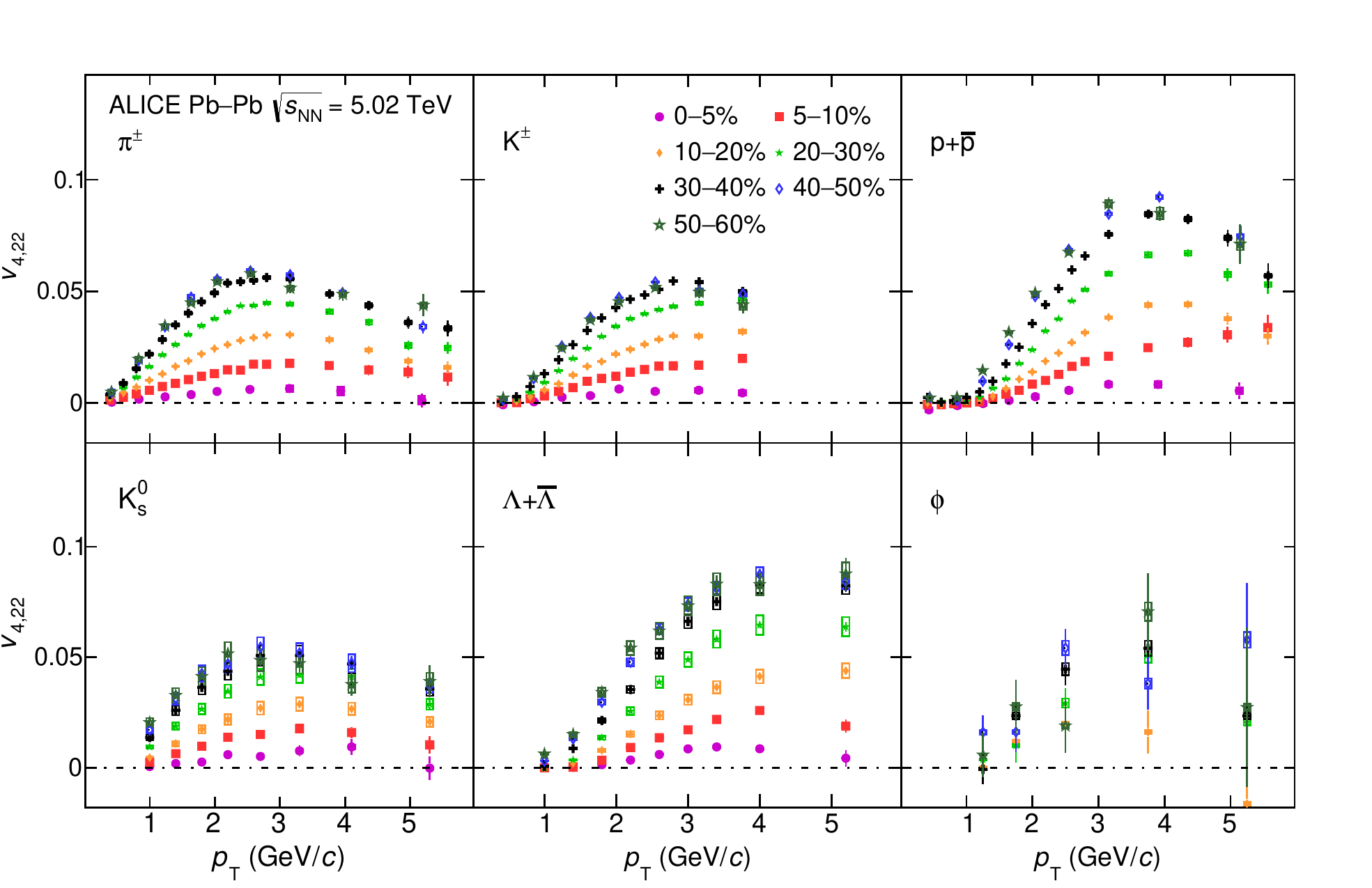}
\end{center}
\caption{The \pT-differential $v_{4,22}$ for different centrality intervals of Pb--Pb collisions at \sNN~grouped by particle species. Statistical and systematic uncertainties are shown as bars and boxes, respectively.}
\label{v422_centralityDependence}
\end{figure}

Figure \ref{v422_centralityDependence} presents the magnitude of the non-linear mode for the fourth order flow coefficient, $v_{4,22}(p_{\rm{T}})$, for \pion, \kaon, \Ks, \proton, \lambdas~and the $\phi$-meson in a wide range of centrality intervals, i.e.\,0--5\% up to 50--60\%. For the $\phi$-meson, the results are reported from the 10--20\% up to the 40--50\% centrality interval, where $v_{4,22}$ can be measured accurately. The magnitude of this non-linear flow mode rises steeply with increasing centrality interval from 0--5\% to 40--50\% for all particle species. This increase is expected as $v_{4,22}$ reflects the contribution of the second order eccentricity, $\varepsilon_{2}$, which increases from central to peripheral collisions, in $v_{4}$ \cite{Alver:2010gr, Acharya:2017zfg}. For more peripheral collisions (i.e. 50--60\%), the magnitude of $v_{4,22}$ does not increase further with respect to the neighbouring centrality interval (40--50\%). This effect that was observed also in $v_n$ measurements \cite{Abelev:2014pua,Acharya:2018zuq} is probably due to the shorter lifetime of the produced system in more peripheral collisions, which prevents $v_{4,22}$ from developing further.

\begin{figure}[!htb]
\begin{center}
\includegraphics[scale=0.82]{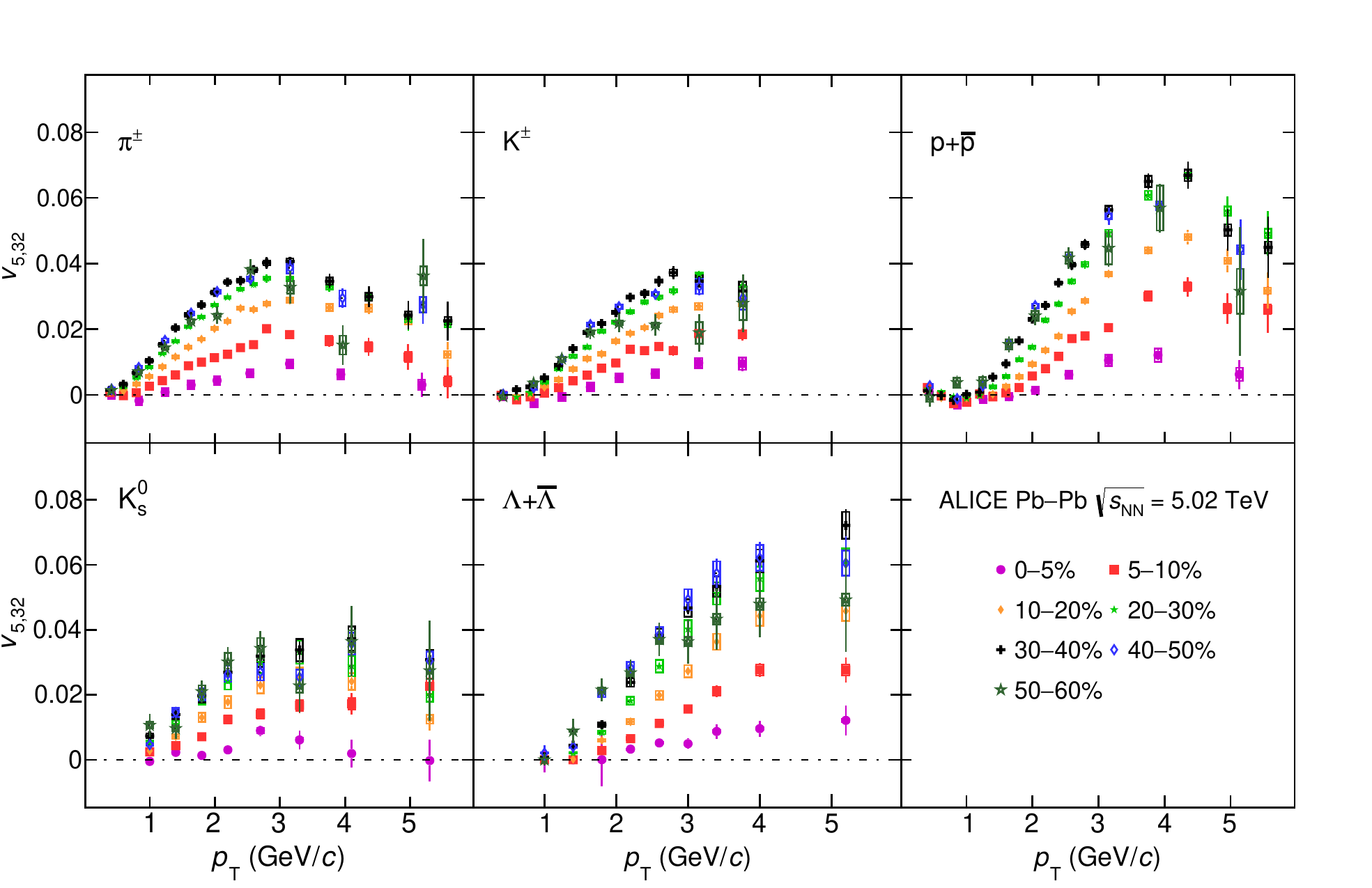}
\end{center}
\caption{The \pT-differential $v_{5,32}$ for different centrality intervals of Pb--Pb collisions at \sNN~grouped by particle species. Statistical and systematic uncertainties are shown as bars and boxes, respectively.}
\label{v523_centralityDependence}
\end{figure}

Figure \ref{v523_centralityDependence} presents the non-linear mode for the fifth order flow coefficient, i.e. $v_{5,32}(p_{\rm{T}})$, of \pion, \kaon, \Ks, \proton, and \lambdas~for the same range of centrality intervals, i.e. 0--5\% up to 50--60\%. Statistical precision limits extending the measurements of non-linear flow modes of the $\phi$-meson for ${\rm n}>4$. The measurements show a significant increase in the magnitude of this non-linear flow mode with increasing centrality percentile. This is due to the fact that $v_{5,32}(p_{\rm{T}})$ has a contribution from both $\varepsilon_{2}$ and $\varepsilon_{3}$. It is shown in MC studies that $\varepsilon_{2}$ and to a smaller extent, $\varepsilon_{3}$ increase for peripheral collisions \cite{Alver:2010gr}. 

\begin{figure}[!htb]
\begin{center}
\includegraphics[scale=0.82]{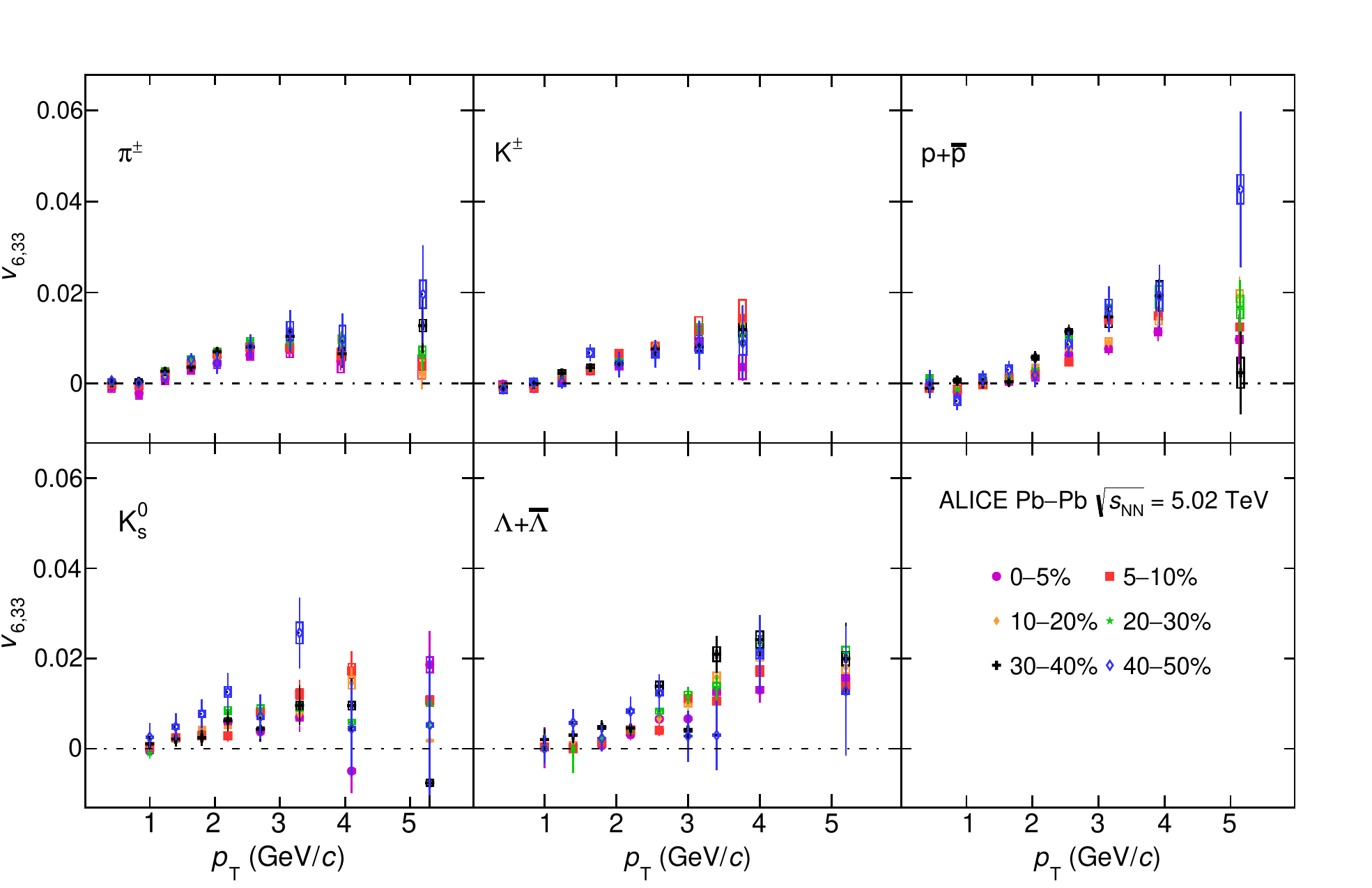}
\end{center}
\caption{The \pT-differential $v_{6,33}$ for different centrality intervals of Pb--Pb collisions at \sNN~grouped by particle species. Statistical and systematic uncertainties are shown as bars and boxes, respectively.}
\label{v633_centralityDependence}
\end{figure}

\begin{figure}[!htb]
\begin{center}
\includegraphics[scale=0.82]{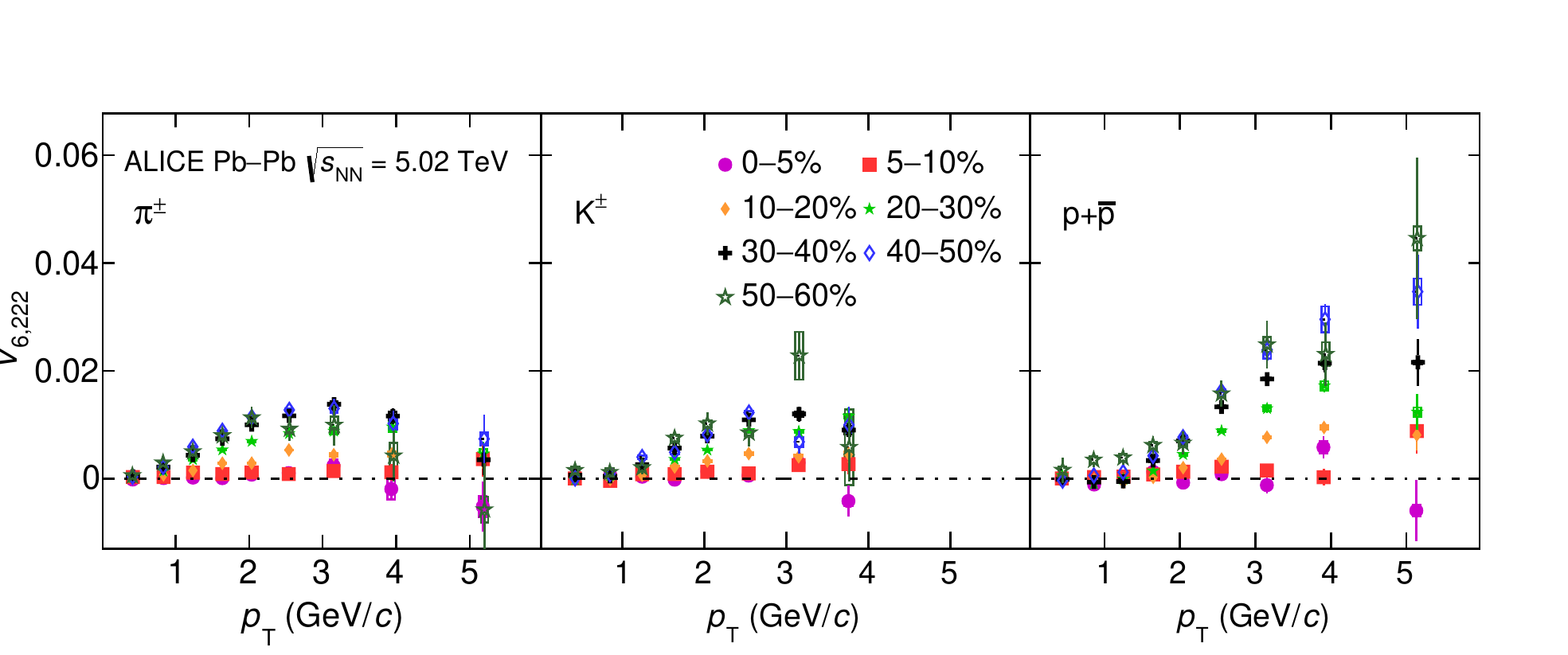}
\end{center}
\caption{The \pT-differential $v_{6,222}$ for different centrality intervals of Pb--Pb collisions at \sNN~grouped by particle species. Statistical and systematic uncertainties are shown as bars and boxes, respectively.}
\label{v6222_centralityDependence}
\end{figure}

Figures \ref{v633_centralityDependence} and \ref{v6222_centralityDependence} present the non-linear terms for the sixth order flow coefficient, i.e. $v_{6,33}(p_{\rm{T}})$ for \pion, \kaon, \Ks, \proton~and \lambdas~for the 0--5\% up to 40--50\% centrality intervals and $v_{6,222}(p_{\rm{T}})$ for \pion, \kaon, \proton~for the 0--5\% up to 50--60\% centrality intervals. As expected, measurements of $v_{6,222}(p_{\rm{T}})$ which probe the contribution of $\varepsilon_2$, show an increase in the magnitude of this non-linear flow mode with increasing centrality percentile. On the other hand, the $v_{6,33}(p_{\rm{T}})$ measurements, which probe the contribution of $\varepsilon_3$, present little to no dependence on centrality as previously observed for charged particles in \cite{Acharya:2017zfg}. 

\newpage

\begin{figure}[!htb]
\begin{center}
\includegraphics[scale=0.82]{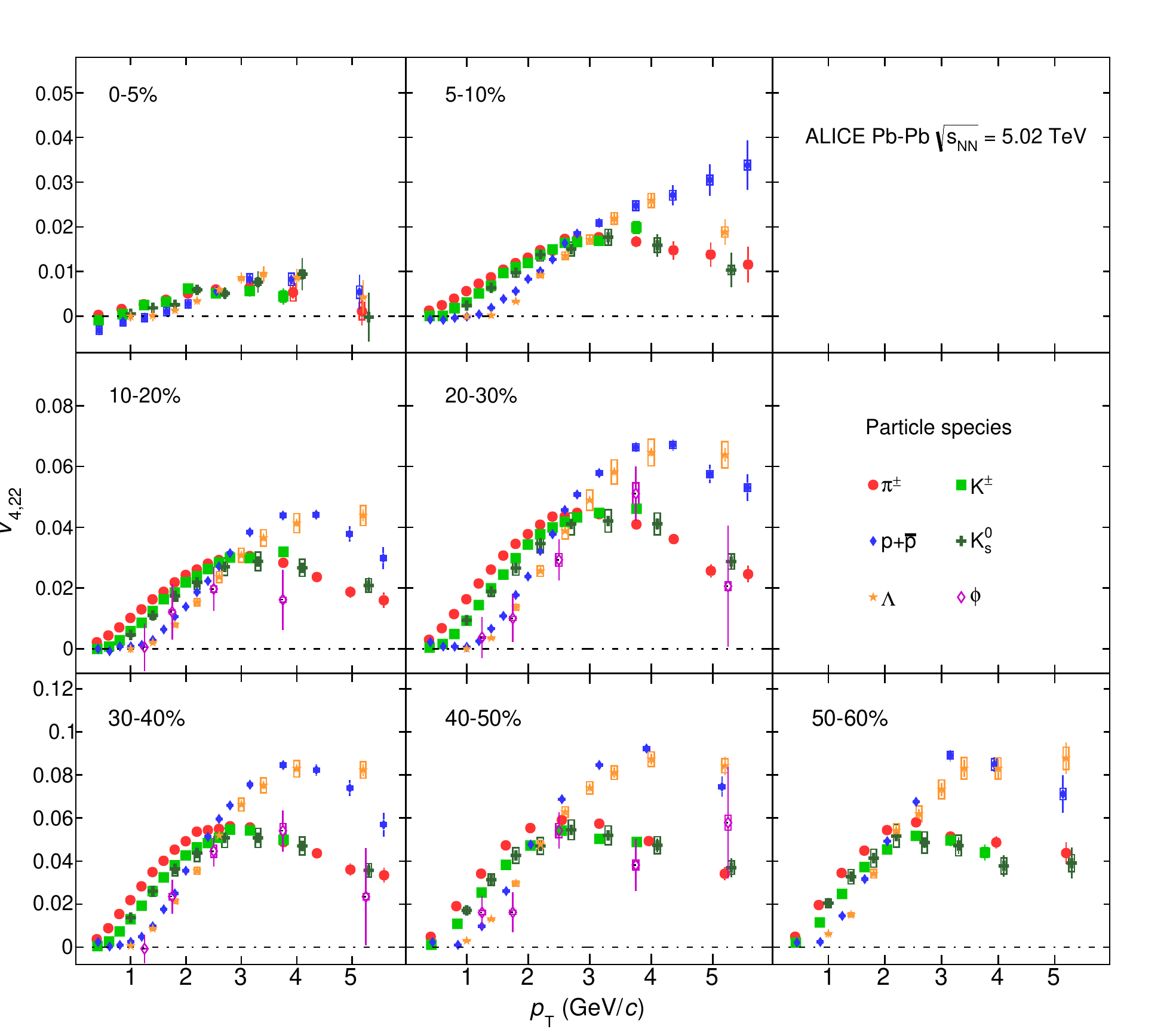}
\end{center}
\caption{The \pT-differential $v_{4,22}$ for different particle species grouped into different centrality intervals of Pb--Pb collisions at \sNN. Statistical and systematic uncertainties are shown as bars and boxes, respectively.}
\label{v422_particleDependence}
\end{figure}

\begin{figure}[!htb]
\begin{center}
\includegraphics[scale=0.82]{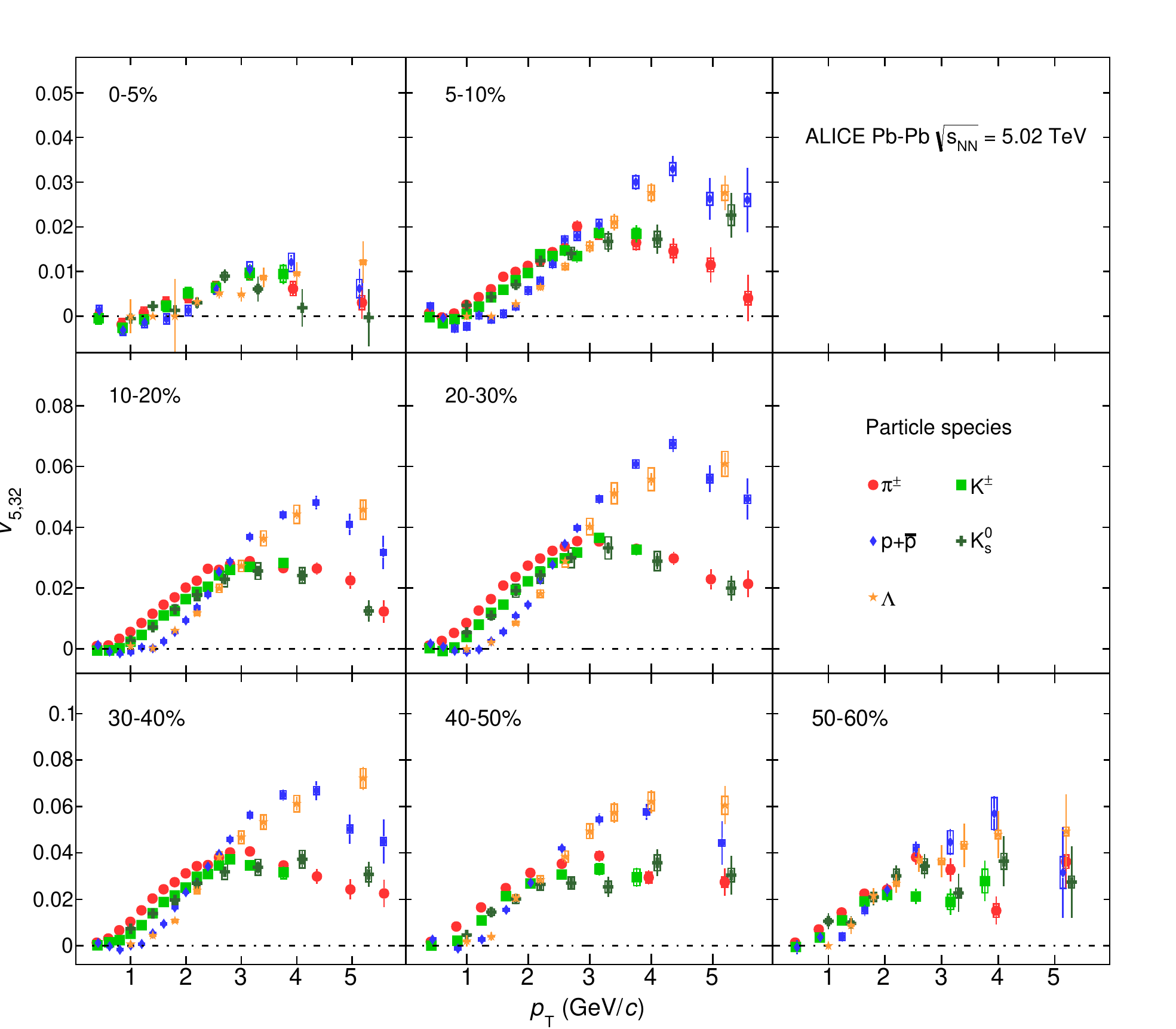}
\end{center}
\caption{The \pT-differential $v_{5,32}$ for different particle species grouped into different centrality intervals of Pb--Pb collisions at \sNN. Statistical and systematic uncertainties are shown as bars and boxes, respectively.}
\label{v523_particleDependence}
\end{figure}

\begin{figure}[!htb]
\begin{center}
\includegraphics[scale=0.82]{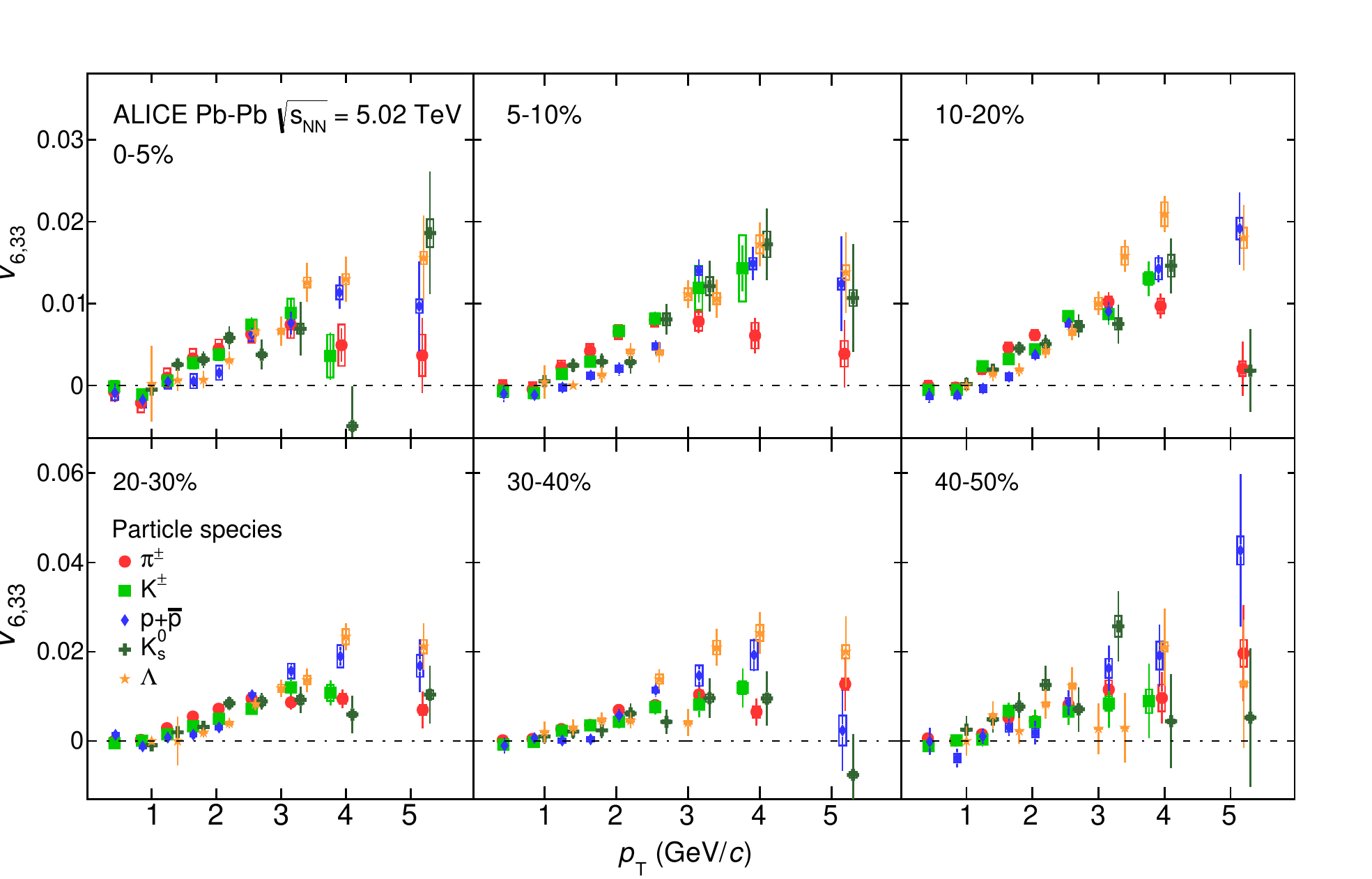}

\end{center}
\caption{The \pT-differential $v_{6,33}$ for different particle species grouped into different centrality intervals of Pb--Pb collisions at \sNN. Statistical and systematic uncertainties are shown as bars and boxes, respectively.}
\label{v633_particleDependence}
\end{figure}

\begin{figure}[!htb]
\begin{center}
\includegraphics[scale=0.82]{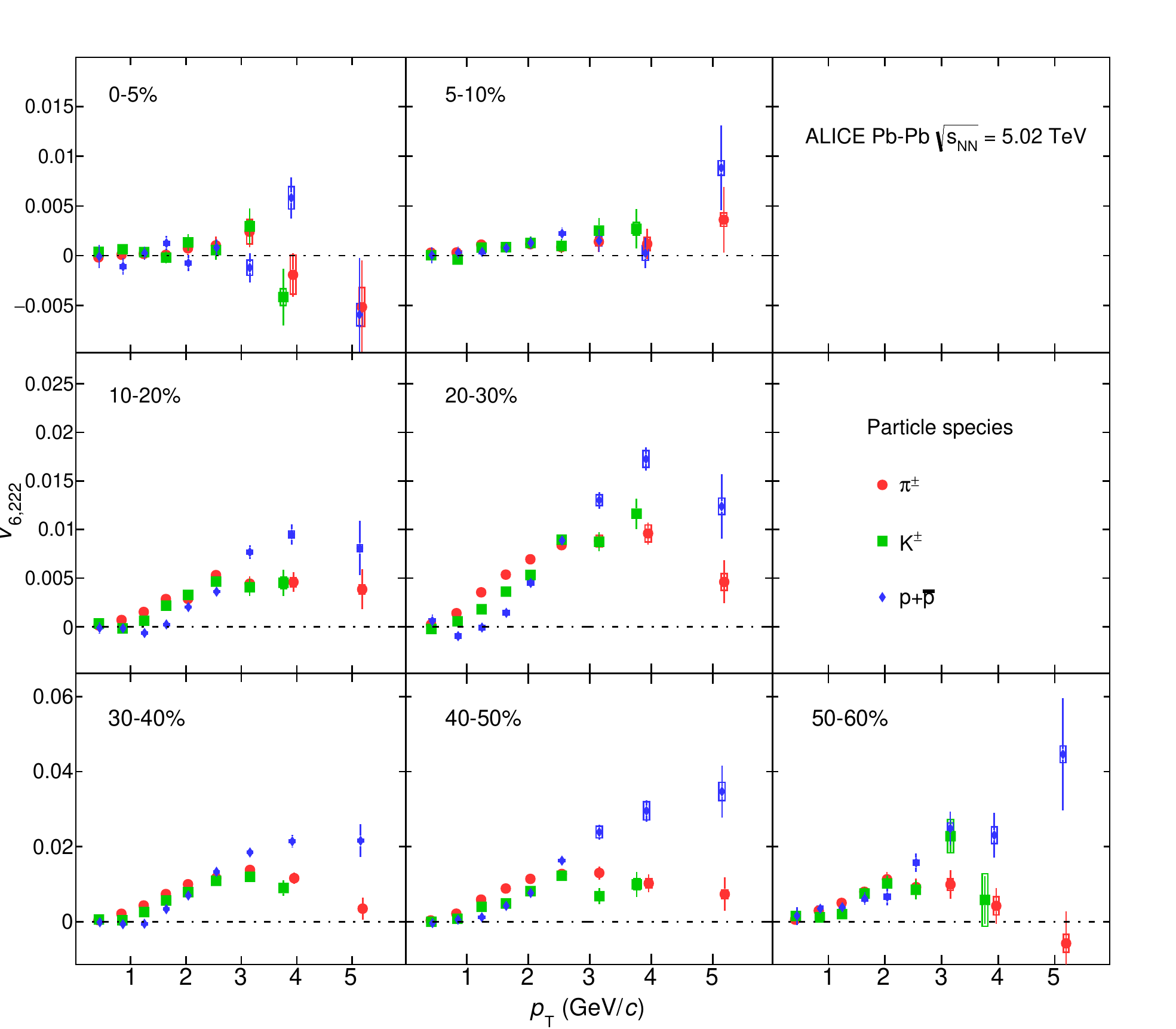}

\end{center}
\caption{The \pT-differential $v_{6,222}$ for different particle species grouped into different centrality intervals of Pb--Pb collisions at \sNN. Statistical and systematic uncertainties are shown as bars and boxes, respectively.}
\label{v6222_particleDependence}
\end{figure}

In Fig. \ref{v422_particleDependence} the same data points are grouped by centrality interval to highlight how $v_{4,22}$ develops for a given centrality for various particle species as a function of \pT.
A clear mass ordering can be seen in the low \pT~region (i.e. \pT $< 2.5$ \GeV) for all collision centralities. This mass ordering arises from the interplay between radial flow and the initial spatial anisotropy, generated from both the geometry and the fluctuating initial energy density profile. This creates a depletion in the particle spectra at lower \pT~values which becomes larger in-plane than out-of plane due to the velocity profile. This naturally leads to lower $v_{4,22}$(\pT) values for heavier particles \cite{Voloshin:1996nv, Huovinen:2001cy, Shen:2011eg}. Similarly, Figs. \ref{v523_particleDependence}, \ref{v633_particleDependence} and \ref{v6222_particleDependence} show the \pT-differential $v_{5,32}$, $v_{6,33}$ and $v_{6,222}$, respectively, of different particle species for each centrality interval. A clear mass ordering is seen in the low \pT~region, (i.e. \pT $< 2.5$ \GeV), for $v_{5,32}(p_{\rm{T}})$ and to a smaller extent for $v_{6,33}(p_{\rm{T}})$ as well as for some centrality intervals of $v_{6,222}(p_{\rm{T}})$.

\newpage

In addition, in the intermediate \pT~region (for \pT $> 2.5$ \GeV) the data points of Figs. \ref{v422_particleDependence}-\ref{v6222_particleDependence} exhibit a particle type grouping. In particular, the data points form two groups, one for mesons and one for baryons with the values of $v_{n,mk}$ of the latter being larger. This particle type grouping was previously observed in $v_{n}$ measurements of various particle species \cite{Abelev:2014pua,Adam:2016nfo,Acharya:2018zuq,Adams:2003am,Abelev:2007qg,Adler:2003kt,Adare:2006ti}. 
This grouping was explained in Ref. \cite{Molnar:2003ff} in the picture of particle production via quark coalescence indicating that flow develops at the partonic stage. In this picture, known as NCQ scaling, the flow of mesons (baryons) is roughly twice (thrice) the flow of their constituent quarks in the intermediate transverse momentum region \cite{Voloshin:2002wa,Molnar:2003ff}. The ALICE measurements show that this scaling at the LHC energies holds at an approximate level of 20\% for $v_{n}$ \cite{Abelev:2014pua,Adam:2016nfo,Acharya:2018zuq}. 

\begin{figure}[!htb]
\begin{center}
\includegraphics[scale=0.82]{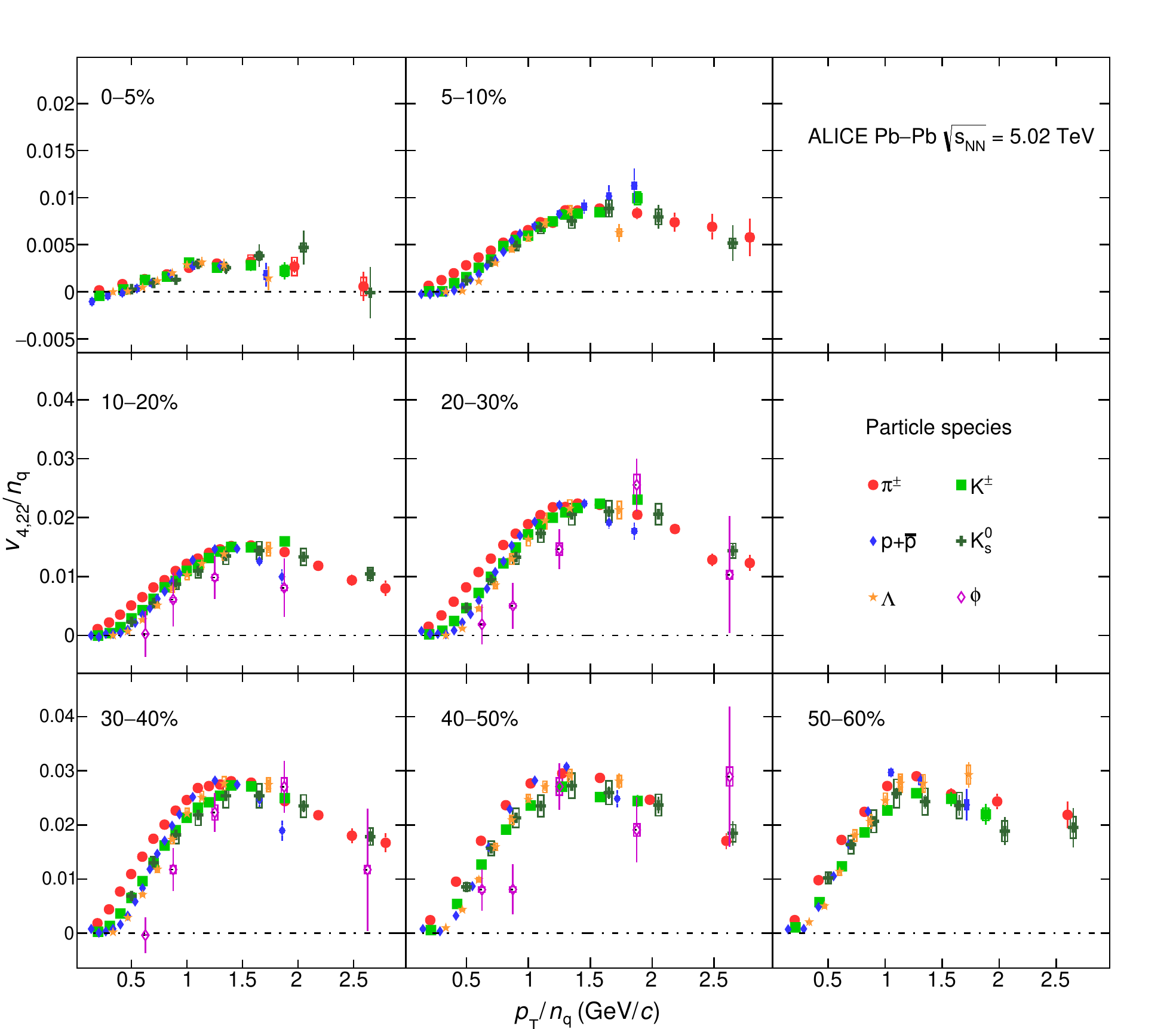}
\end{center}
\caption{The $p_{\rm{T}}/n_{q}$-dependence of $v_{4,22}/n_{q}$ for different particle species grouped into different centrality intervals of Pb--Pb collisions at \sNN. Statistical and systematic uncertainties are shown as bars and boxes, respectively.}
\label{v422_NCQ}
\end{figure}

\begin{figure}[!htb]
\begin{center}
\includegraphics[scale=0.82]{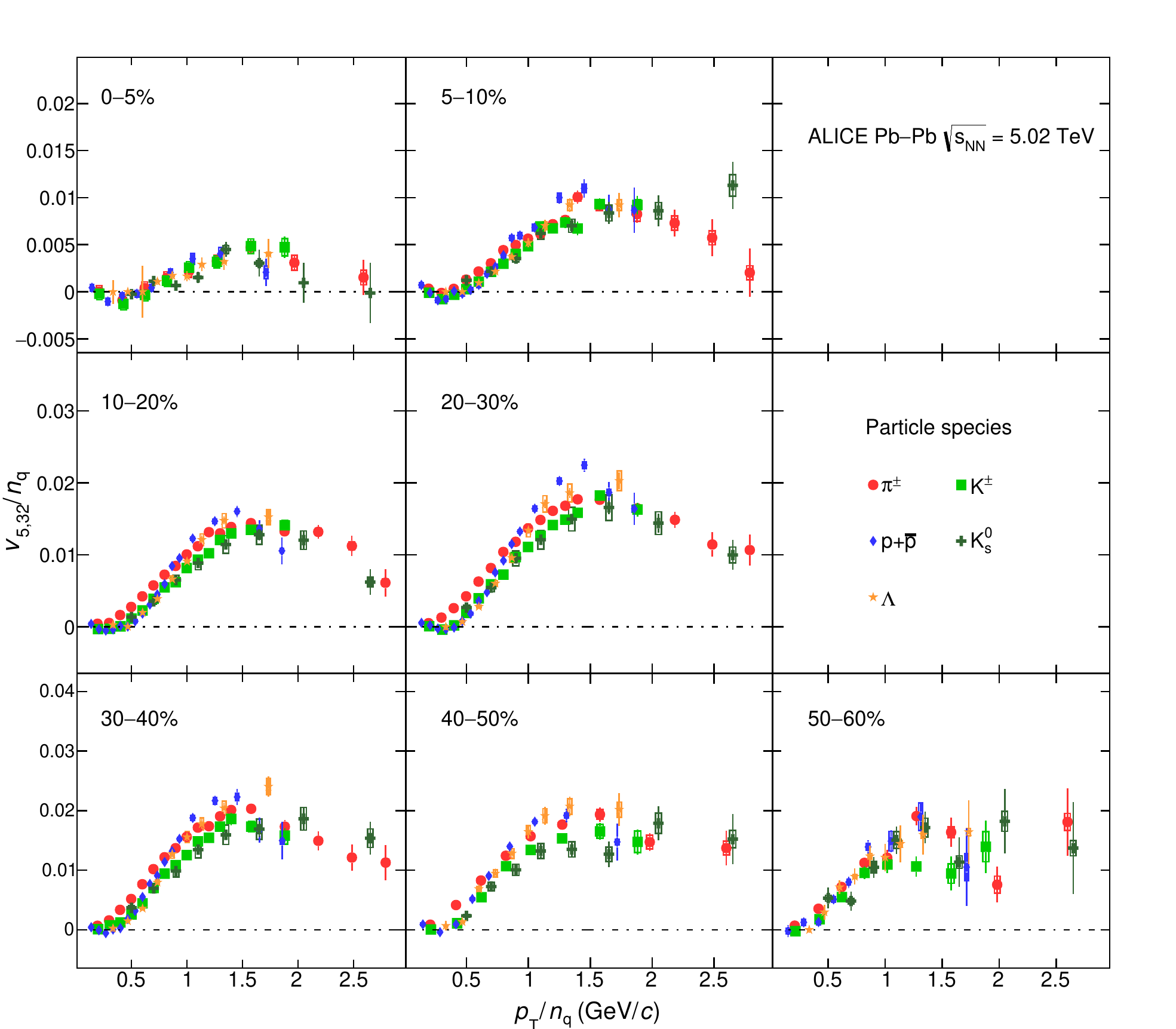}
\end{center}
\caption{The $p_{\rm{T}}/n_{q}$-dependence of $v_{5,32}/n_{q}$ for different particle species grouped into different centrality intervals of Pb--Pb collisions at \sNN. Statistical and systematic uncertainties are shown as bars and boxes, respectively.}
\label{v523_NCQ}
\end{figure}

\begin{figure}[!htb]
\begin{center}
\includegraphics[scale=0.82]{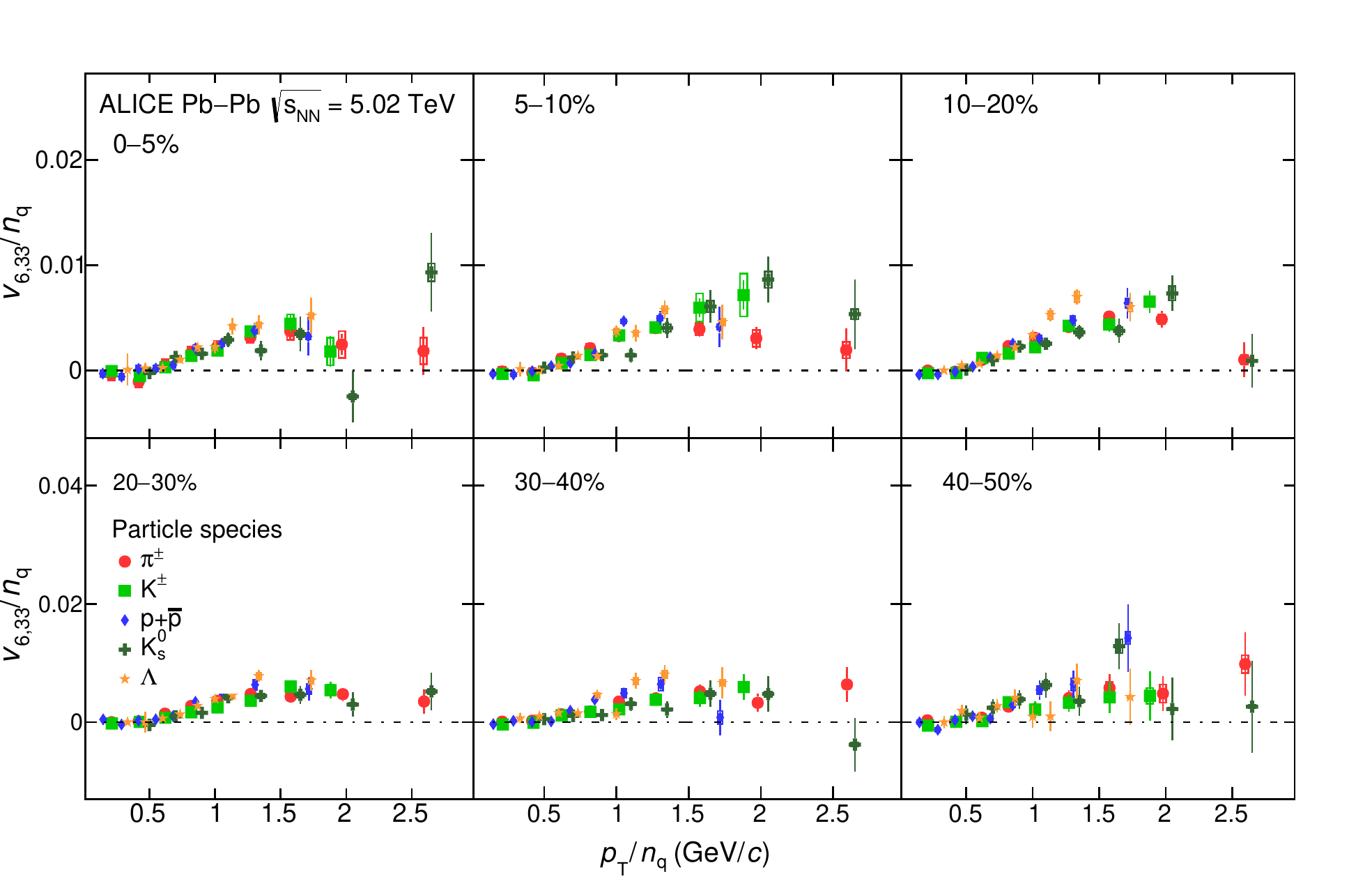}
\end{center}
\caption{The $p_{\rm{T}}/n_{q}$-dependence of $v_{6,33}/n_{q}$ for different particle species grouped into different centrality intervals of Pb--Pb collisions at \sNN. Statistical and systematic uncertainties are shown as bars and boxes, respectively.}
\label{v633_NCQ}
\end{figure}

\begin{figure}[!htb]
\begin{center}
\includegraphics[scale=0.82]{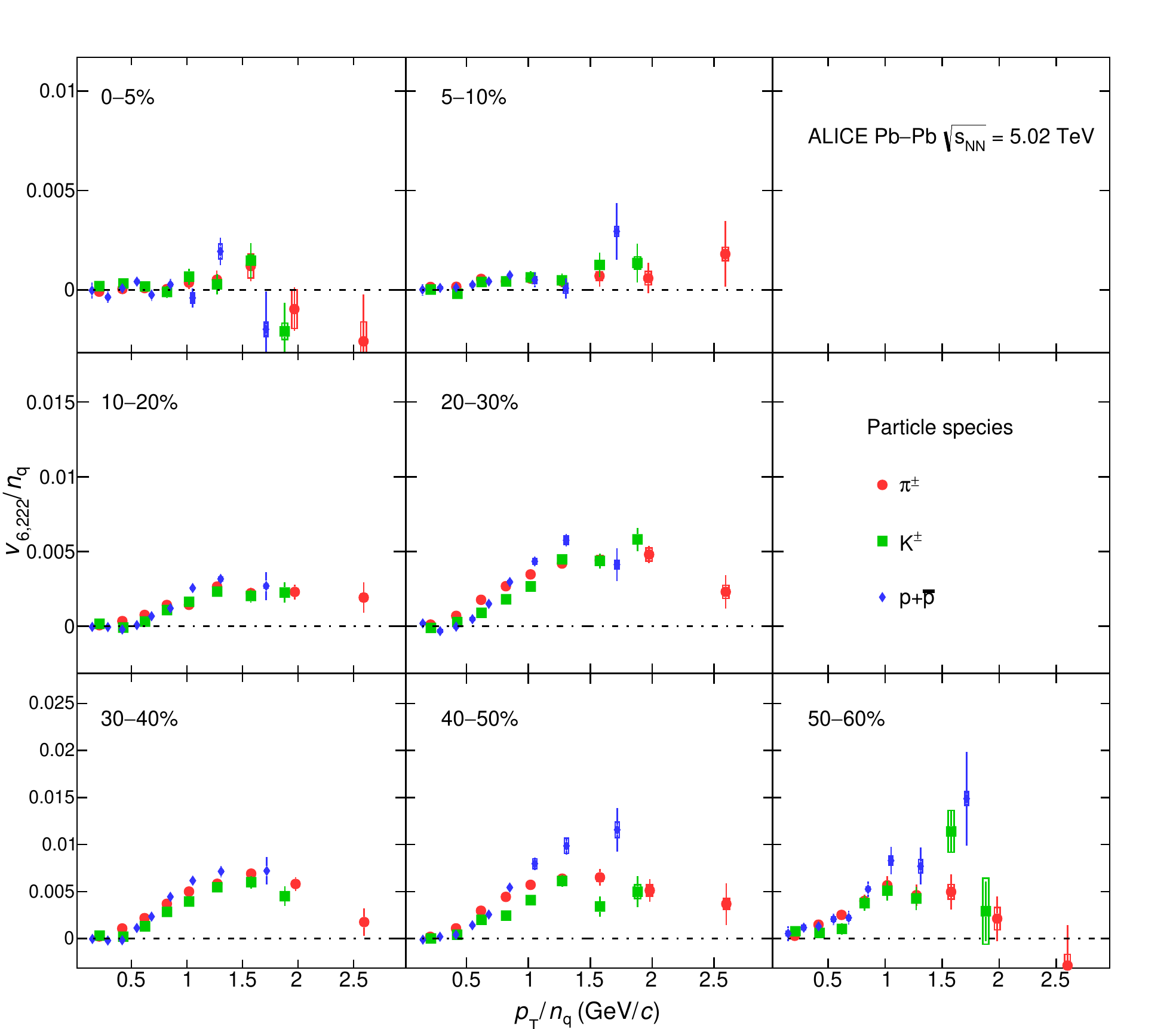}
\end{center}
\caption{The $p_{\rm{T}}/n_{q}$-dependence of $v_{6,222}/n_{q}$ for different particle species grouped into different centrality intervals of Pb--Pb collisions at \sNN. Statistical and systematic uncertainties are shown as bars and boxes, respectively.}
\label{v6222_NCQ}
\end{figure}

Figures \ref{v422_NCQ}, \ref{v523_NCQ}, \ref{v633_NCQ} and \ref{v6222_NCQ} present $v_{4,22}$, $v_{5,32}$, $v_{6,33}$ and $v_{6,222}$, respectively, scaled by the number of constituent quarks ($n_{q}$) as a function of \pTnq~for \pion, \kaon, \Ks, \proton, \lambdas~and the $\phi$-meson grouped in different centrality intervals. The scaling is consistent with the observations reported for higher order anisotropic flow coefficients \cite{Acharya:2018zuq}. It is seen that for the non-linear flow modes this scaling holds at an approximate level ($\pm$20\%) for \pT $> 1$ \GeVc,~where quark coalescence is expected to be the dominant process.

\subsection{Comparison with $v_{\rm n}$ of identified particles}
\label{SubSec:comparewithvn}

\begin{figure}[!htb]
\begin{center}
\includegraphics[scale=0.6]{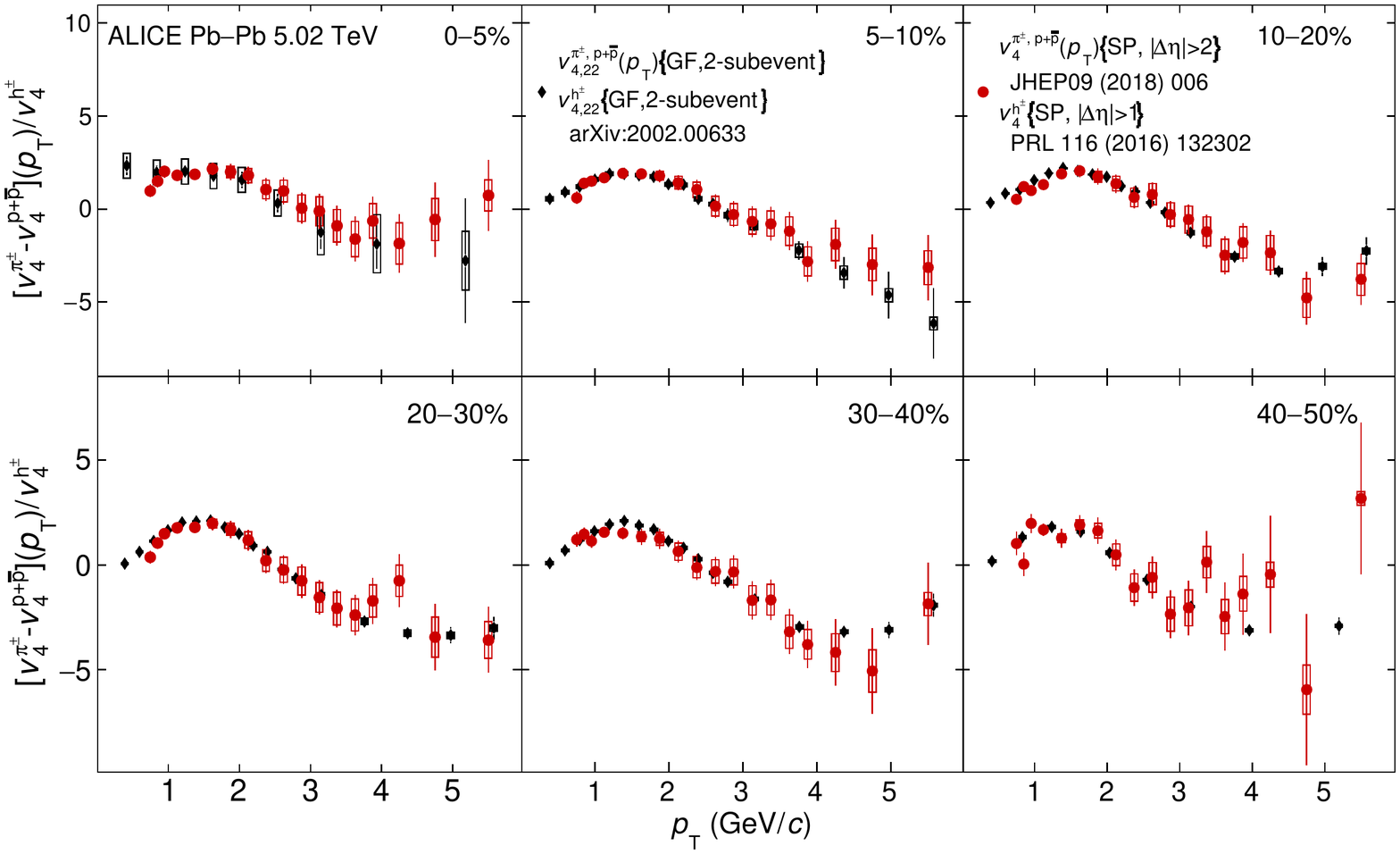}
\end{center}
\caption{The comparison between $[v_{4,22}^{\pi^{\pm}} -  v_{4,22}^{\rm p+\bar{p}}](p_{\rm T}) / v_{4,22}^{\rm h^{\pm}} $ and $[v_{4}^{\pi^{\pm}} -  v_{4}^{\rm p+\bar{p}}](p_{\rm T}) / v_{4}^{\rm h^{\pm}} $  grouped into different centrality intervals of Pb--Pb collisions at \sNN. Statistical and systematic uncertainties are shown as bars and boxes, respectively.}
\label{massOrderingComparison}
\end{figure}

The comparison of the features discussed before i.e. mass ordering and particle type grouping between the non-linear and the anisotropic flow coefficient is of particular interest.  Based on a naive expectation the mass ordering should develop quantitatively in a different way between the non-linear (i.e. due to the dependence on $\varepsilon_{2}^{2}$) and the anisotropic flow coefficient. In parallel, if coalescence is the dominant particle production mechanism in the intermediate \pT~region, one expects a similar grouping between $v_{\rm n}^{\rm NL}$ and $v_{\rm n}$. Such a comparison could only be performed for $v_{4,22}$(\pT) (this study) and the $v_{4}$(\pT) measurements \cite{Acharya:2018zuq} and was done by taking the difference between pions and protons at a given \pT~in both modes and normalising it by the integrated flow of the corresponding mode for charged particles \cite{Adam:2016izf} ($[v_{4}^{\pi^{\pm}} -  v_{4}^{\rm p+\bar{p}}](p_{\rm T}) / v_{4}^{\rm h^{\pm}}$). This comparison is shown in Fig. \ref{massOrderingComparison} for the 0--5\% up to the 40--50\% centrality interval. It can be seen that in the low \pT~region (\pT~< $2.5-3$ \GeV) where the mass ordering is prominent, the data points exhibit a general agreement for all centrality intervals. However, there is a hint that the relative ratio for $v_{4,22}$ is smaller than the one of the $v_{4}$ for \pT~$< 0.8$ \GeV~and for the centrality intervals 0--30\%.
 If this difference and its centrality dependence persist for low values of \pT, it could indicate that the hydrodynamic evolution is reflected differently in $v_{4}$ and $v_{4,22}$ and could be explained by the contribution of $\varepsilon_{2}^{2}$.  As stated earlier, the mass splitting is a result of an interplay of radial and anisotropic flow, leading to a stronger in-plane expansion compared to out-of-plane, and the particle thermal motion. Particles with larger mass have smaller thermal velocities, and are thus affected stronger by the difference between in- and out-of-plane expansion velocities, thus leading to the mass splitting of $v_{n}$(\pT). The comparison of the \pT~dependence of $v_{n}^{\rm NL}$ and $v_{n}$ can therefore provide a unique opportunity to test this picture, as it would allow results for the cases of exactly the same average radial flow and temperature, but differing in anisotropic flow to be compared. On the other hand, in the intermediate \pT~region (\pT~> 2.5 \GeV), the same comparison shows that the results are compatible in all centrality intervals within one standard deviation. This implies a similar particle type grouping between $v_{4}$ and $v_{4,22}$ which is in line with the expectation that quark coalescence affects both flow modes  similarly.

\newpage
\subsection{Comparison with models}
\label{SubSec:hydro}

The comparison of various anisotropic flow measurements and hydrodynamic calculations are presented and discussed in great detail in \cite{Xu:2016hmp, McDonald:2016vlt, Zhao:2017yhj}. A recent comparison between $v_{n}$ measurements reported by the ALICE collaboration \cite{Acharya:2018zuq} and two hydrodynamic calculations from \cite{Zhao:2017yhj} shed new light on the initial conditions and the transport properties of the created system in Pb--Pb collisions. Both hydrodynamic calculations are based on iEBE-VISHNU \cite{Shen:2014vra}, an event-by-event version of the VISHNU hybrid model \cite{Song:2010aq} coupling $2+1$~dimensional viscous hydrodynamics (VISH2+1) \cite{Song:2007fn} to a hadronic cascade model (UrQMD). The initial conditions used for these calculations are described by AMPT \cite{Lin:2004en} and \trento~ \cite{Moreland:2014oya}, both with $\tau_{0}$=0.6 fm/$c$ and $T_{sw}$ =148 MeV \cite{Bernhard:2016tnd}. For AMPT initial conditions, constant values of specific shear viscosity over entropy density ($\eta/s =0.08$, the lower limit conjectured by AdS/CFT) and bulk viscosity over entropy density ($\zeta/s = 0$) are utilised. The version of the model that uses \trento~ \cite{Moreland:2014oya} initial conditions incorporates temperature dependent specific shear and bulk viscosities extracted from the global bayesian analysis \cite{Bernhard:2016tnd}. \footnote{ For simplicity in the rest of this article the model with AMPT initial conditions, $\eta/s =0.08$ and $\zeta/s =0$ is referred to as AMPT and the model with \trento~initial conditions, $\eta/s(\rm{T})$ and $\zeta/s(\rm{T})$ is referred to as \trento.} 

The comparison between $v_{n}$ measurements and these two hydrodynamic calculations illustrates a qualitative agreement. This agreement between the data and the models depends on the particle species, transverse momentum range and centrality percentile. Overall, the AMPT model reproduces the measurements more accurately than the \trento~model \cite{Acharya:2018zuq}. In order to further investigate the performance of these two models in reproducing the $v_{\rm n}$ measurements, and provide a quantitative comparison, the relative ratios between each model and the measurements of \pion, \kaon~and \proton~are obtained. Table \ref{ModelDataComparisonflow} summarises these relative ratios. The values represent the ranges across all centralities that each model is able to describe the measurements of $v_n$ for each particle species. Comparisons between the performance of the two models show that the AMPT calculations reproduce $v_{2}$ slightly better that \trento. Both models reproduce the $v_{3}$ measurements relatively better than the $v_{2}$, however AMPT performs better than \trento. Finally, the comparison between the models and the $v_{4}$ measurements show that AMPT has an absolute better performance compared to \trento. These values should be taken with caution as $v_{4}$ has larger uncertainties with respect to $v_{3}$ and $v_{2}$. 

\begin{table}[!h]
\centering
\caption{List of minimum and maximum values of the fit with a constant function to relative ratios between data and each model for  $v_{\rm n} ({\rm n}=2,3,4)$ of \pion, \kaon~and \proton. The percentages show deviations of the fit from unity obtained for the 0--5\% up to 40--50\% centrality intervals.}
\resizebox{0.81\textwidth}{!}{\begin{tabular}{ |p{4.5cm} |l|c|c|c|c|c|c|c|c|c|}
\hline
\multicolumn{1}{| c |}{} & \multicolumn{3}{| c |}{ $v_{2}$ } & \multicolumn{3}{| c |}{ $v_{3}$} & \multicolumn{3}{| c |}{ $v_{4}$}  \\
\hline
Model  & \pion &  \kaon & \proton &  \pion & \kaon & \proton &  \pion &  \kaon & \proton \\ \hline  \hline
AMPT calculations & 3--13\% & 0--16\% & 0--20\% & 0--8\% & 5--12\% & 0--4\%& 6--12\% & 5--12\% & 0--4\%  \\
\trento~ calculations & 6--17\% & 0--19\% & 3--19\% & 2--15\% & 7--22\% & 0--11\% & 7--25\% & 16--28\% & 0--21\% \\
 \hline
\end{tabular}}
\label{ModelDataComparisonflow}
\end{table}

\begin{table}[!h]
\caption{List of minimum and maximum values of the fit with a constant function to relative ratios between the data and each model for  $v_{\rm n,mk}$ of \pion, \kaon~and \proton. The percentages show deviations of the fit from unity obtained for the 0--10\% up to 50--60\% (40--50\% for $v_{6,33}$) centrality intervals.}
\resizebox{\textwidth}{!}{\begin{tabular}{ |p{4.5cm} |l|c|c|c|c|c|c|c|c|c|c|c|c|}
\hline
\multicolumn{1}{| c |}{} & \multicolumn{3}{| c |}{ $v_{4,22}$ } & \multicolumn{3}{| c |}{ $v_{5,32}$} & \multicolumn{3}{| c |}{ $v_{6,33}$} & \multicolumn{3}{| c |}{ $v_{6,222}$} \\
\hline
Model  & \pion &  \kaon & \proton &  \pion & \kaon & \proton &  \pion &  \kaon & \proton &  \pion &  \kaon & \proton \\ \hline  \hline
AMPT calculations &  5--32\% & 2--30\%  & 3--30\% & 3--28\%  & 5--29\% &  1--65\% & 0--46\% & 0--46\% & 0--97\% & 6--52\% & 0--80\%  & 0--118\% \\
\trento~ calculations & 0--30\% & 4--33\% & 0--21\% &  24--49\% & 33--97\% & 12--58\% & 0--43\% & 0--46\% & 0--95\% & 0--20\% & 0--34\% & 0--78\%\\
 \hline
\end{tabular}}
\label{ModelDataComparisonNLflow}
\end{table}

\begin{figure}[h]
\begin{center}
\includegraphics[scale=0.73]{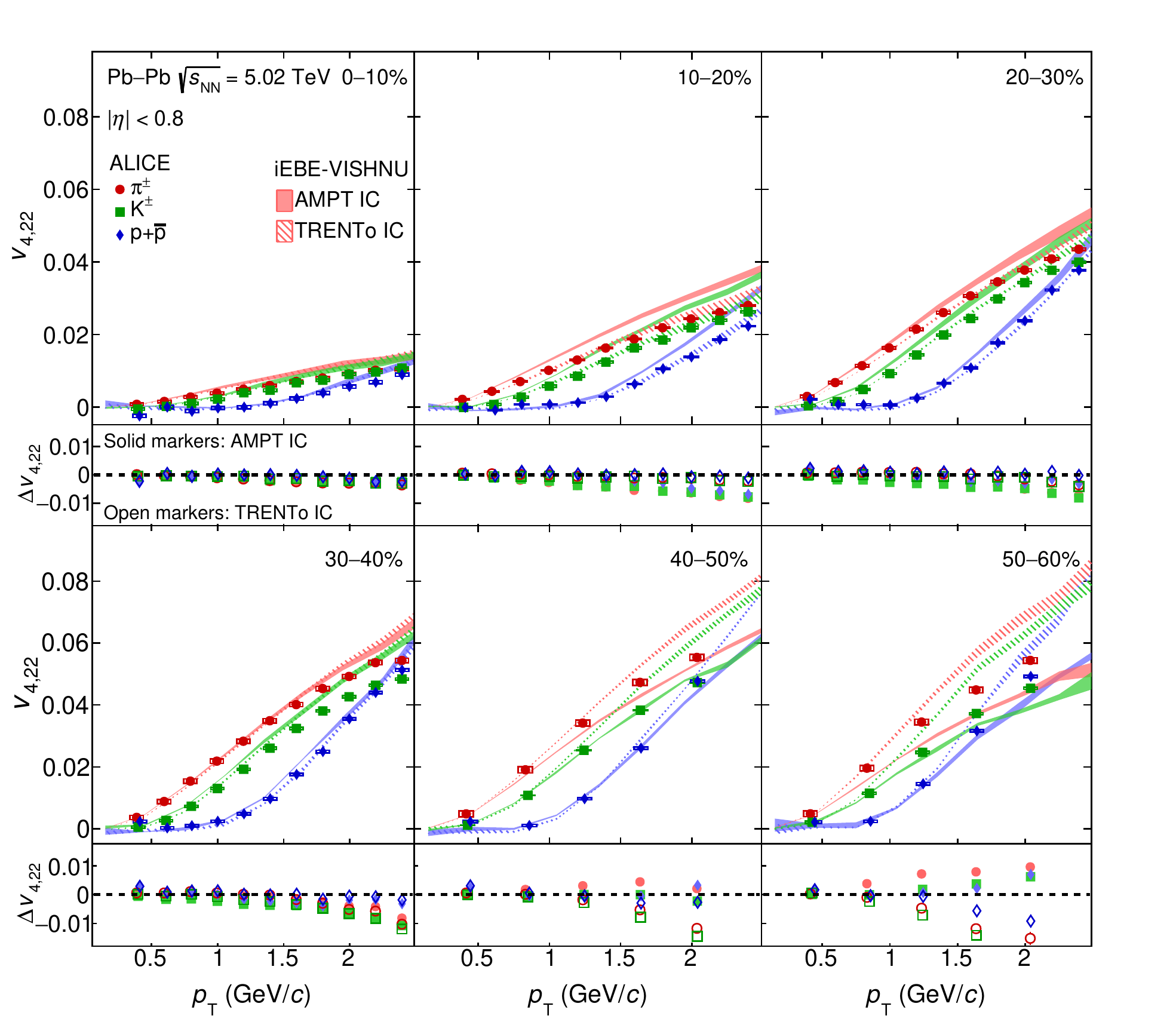}
\end{center}
\caption{The \pT-differential $v_{4,22}$ of \pion, \kaon~and \proton~in the 0--10\% up to 50--60\% centrality intervals of Pb--Pb collisions at \sNN compared with iEBE-VISHNU hybrid models with two different sets of initial parameters: AMPT initial conditions ($\eta/s$= 0.08 and $\zeta/s$ = 0) shown as solid bands and \trento~initial conditions ($\eta/s({\rm T})$ and $\zeta/s({\rm T})$) as hatched bands. The bottom panels show the difference between the measurements and each model. Statistical and systematic uncertainties are shown as bars and boxes, respectively.}
\label{v422_model}
\end{figure}

\begin{figure}[h]
\begin{center}
\includegraphics[scale=0.73]{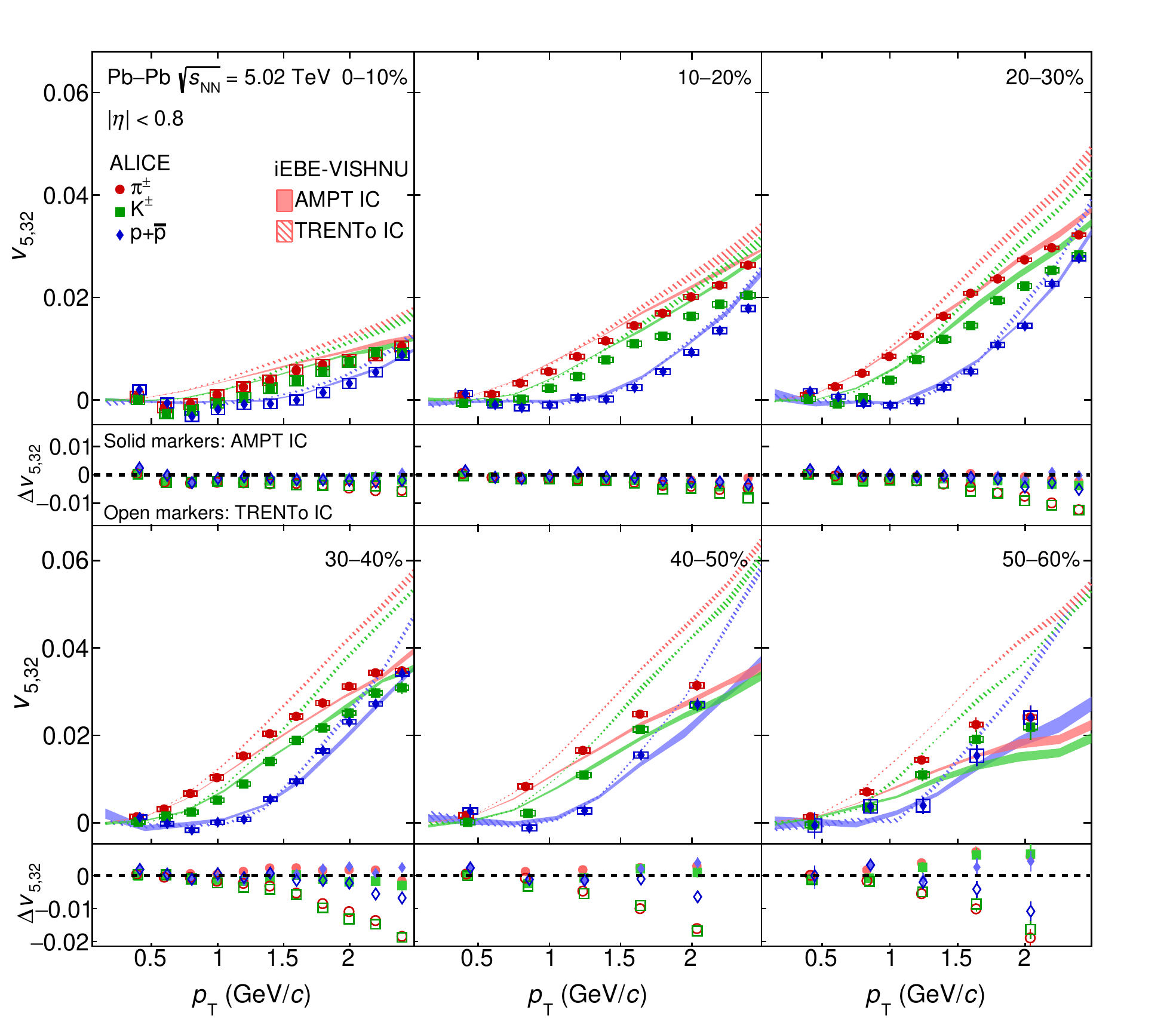}
\end{center}
\caption{The \pT-differential $v_{5,32}$ of \pion, \kaon~and \proton~in the 0--10\% up to 50--60\% centrality intervals of Pb--Pb collisions at \sNN compared with iEBE-VISHNU hybrid models with two different sets of initial parameters: AMPT initial conditions ($\eta/s$= 0.08 and $\zeta/s$ = 0) shown as solid bands and \trento~initial conditions ($\eta/s({\rm T})$ and $\zeta/s({\rm T})$) as hatched bands. The bottom panels show the difference between the measurements and each model. Statistical and systematic uncertainties are shown as bars and boxes, respectively.}
\label{v523_model}
\end{figure}

\begin{figure}[h]
\begin{center}
\includegraphics[scale=0.73]{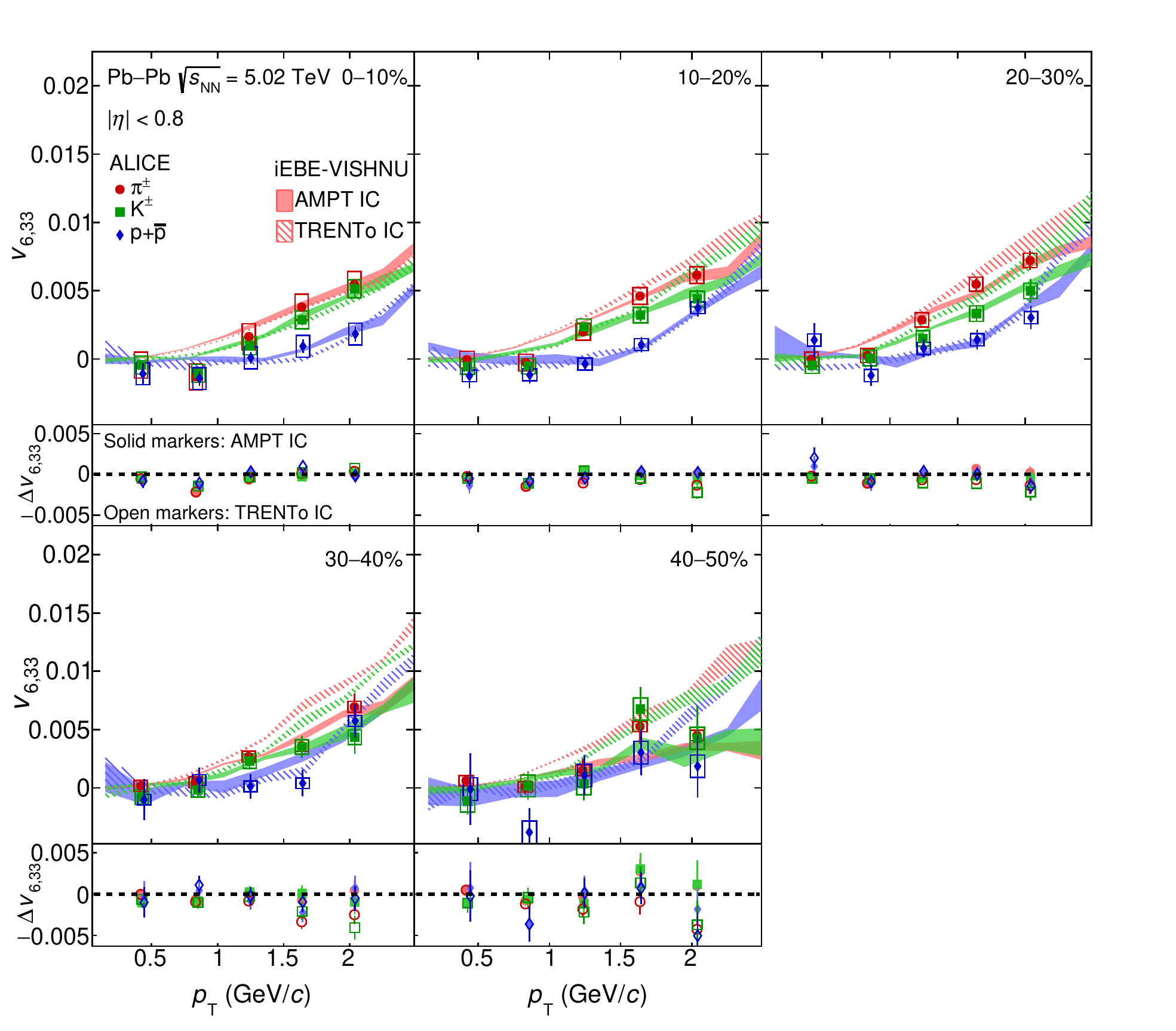}
\end{center}
\caption{The \pT-differential $v_{6,33}$ of \pion, \kaon~and \proton~in the 0--10\% up to 40--50\% centrality intervals of Pb--Pb collisions at \sNN compared with iEBE-VISHNU hybrid models with two different sets of initial parameters: AMPT initial conditions ($\eta/s$= 0.08 and $\zeta/s$ = 0) shown as solid bands and \trento~initial conditions ($\eta/s({\rm T})$ and $\zeta/s({\rm T})$) as hatched bands. The bottom panels show the difference between the measurements and each model. Statistical and systematic uncertainties are shown as bars and boxes, respectively.}
\label{v633_model}
\end{figure}

\begin{figure}[h]
\begin{center}
\includegraphics[scale=0.73]{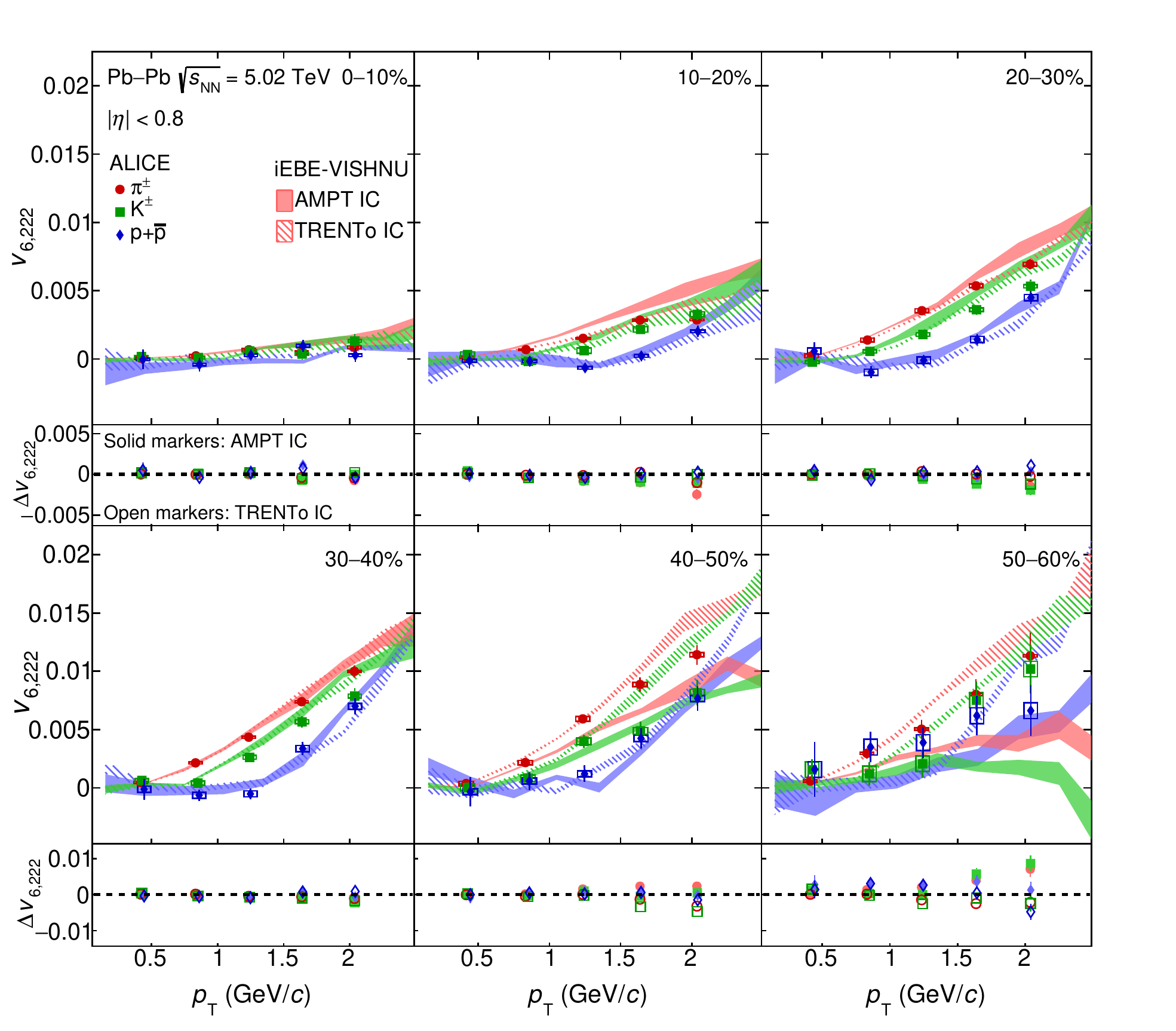}
\end{center}
\caption{The \pT-differential $v_{6,222}$ of \pion, \kaon~and \proton~in the 0--10\% up to 50--60\% centrality intervals of Pb--Pb collisions at \sNN compared with iEBE-VISHNU hybrid models with two different sets of initial parameters: AMPT initial conditions ($\eta/s$= 0.08 and $\zeta/s$ = 0) shown as solid bands and \trento~initial conditions ($\eta/s({\rm T})$ and $\zeta/s({\rm T})$) as hatched bands. The bottom panels show the difference between the measurements and each model. Statistical and systematic uncertainties are shown as bars and boxes, respectively.}
\label{v6222_model}
\end{figure}

\begin{figure}[h]
\begin{center}
\includegraphics[scale=0.73]{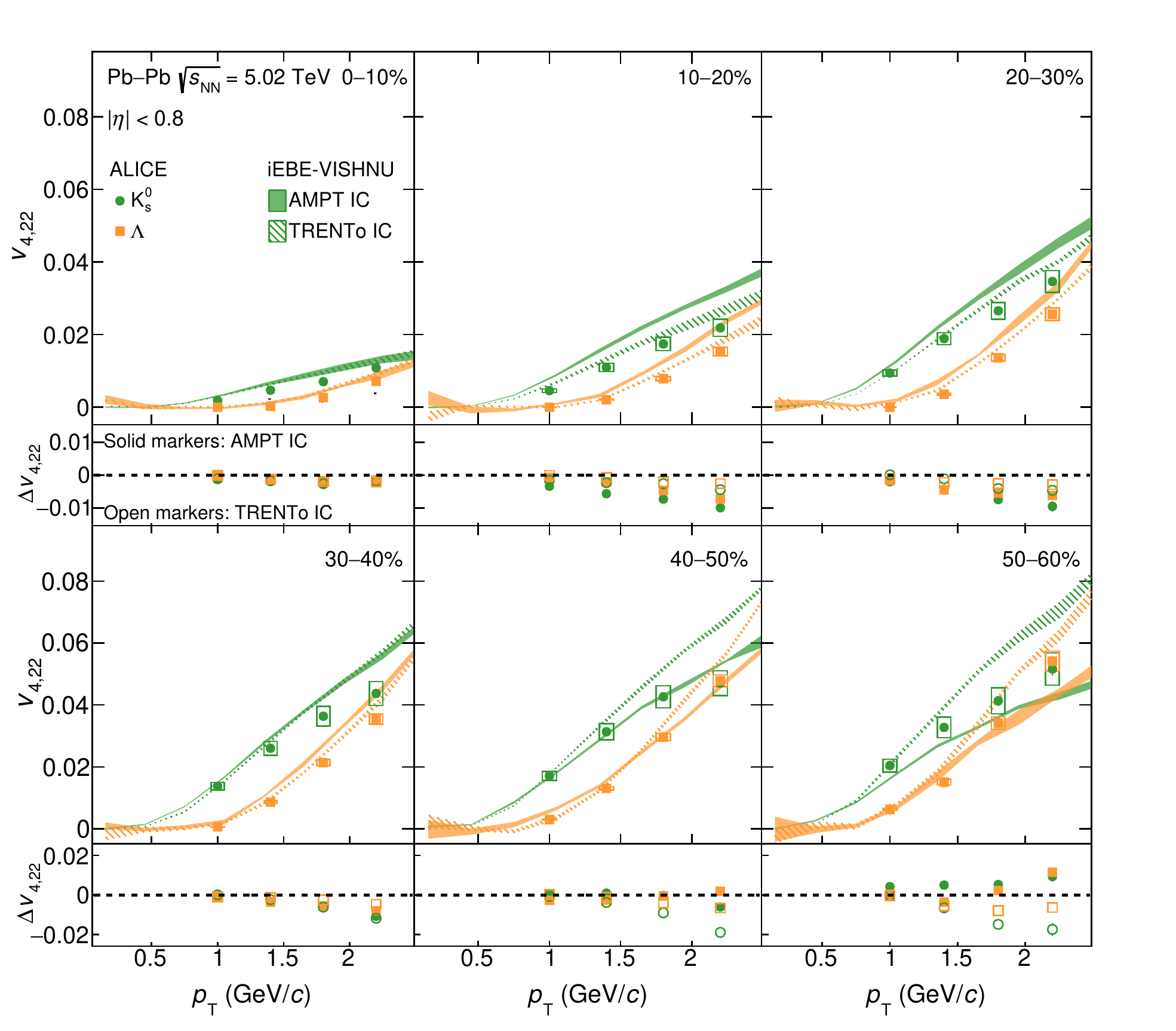}
\end{center}
\caption{The \pT-differential $v_{4,22}$ of \Ks~and \lambdas~in the 0--10\% up to 50--60\% centrality intervals of Pb--Pb collisions at \sNN compared with iEBE-VISHNU hybrid models with two different sets of initial parameters: AMPT initial conditions ($\eta/s$= 0.08 and $\zeta/s$ = 0) shown as solid bands and \trento~initial conditions ($\eta/s({\rm T})$ and $\zeta/s({\rm T})$) as hatched bands. The bottom panels show the difference between the measurements and each model. Statistical and systematic uncertainties are shown as bars and boxes, respectively.}
\label{v422_model_KL}
\end{figure}

\begin{figure}[h]
\begin{center}
\includegraphics[scale=0.73]{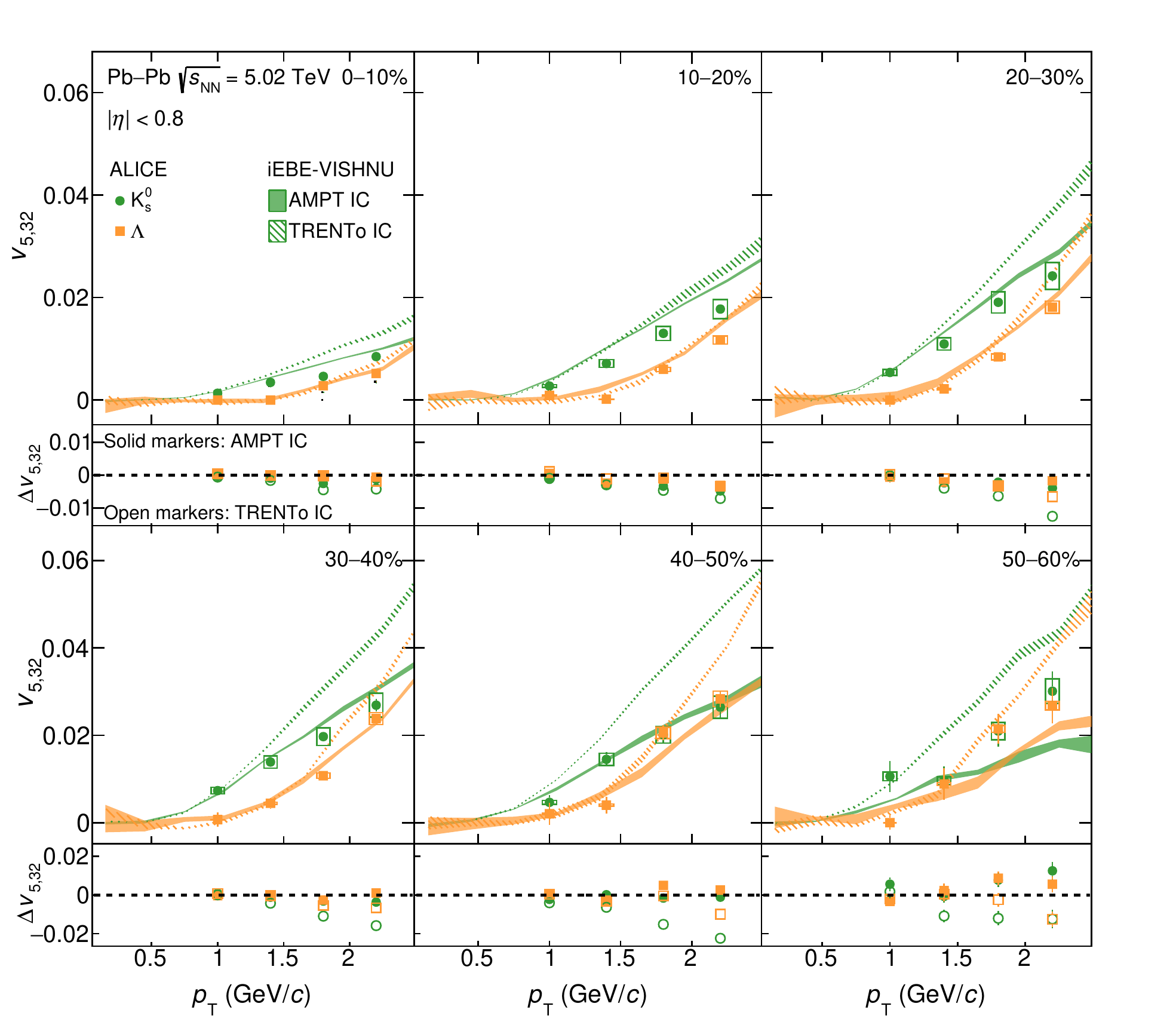}
\end{center}
\caption{The \pT-differential $v_{5,32}$ of \Ks~and \lambdas~in the 0--10\% up to 50--60\% centrality intervals of Pb--Pb collisions at \sNN compared with iEBE-VISHNU hybrid models with two different sets of initial parameters: AMPT initial conditions ($\eta/s$= 0.08 and $\zeta/s$ = 0) shown as solid bands and \trento~initial conditions ($\eta/s({\rm T})$ and $\zeta/s({\rm T})$) as hatched bands. The bottom panels show the difference between the measurements and each model. Statistical and systematic uncertainties are shown as bars and boxes, respectively.}
\label{v523_model_KL}
\end{figure}

\begin{figure}[h]
\begin{center}
\includegraphics[scale=0.73]{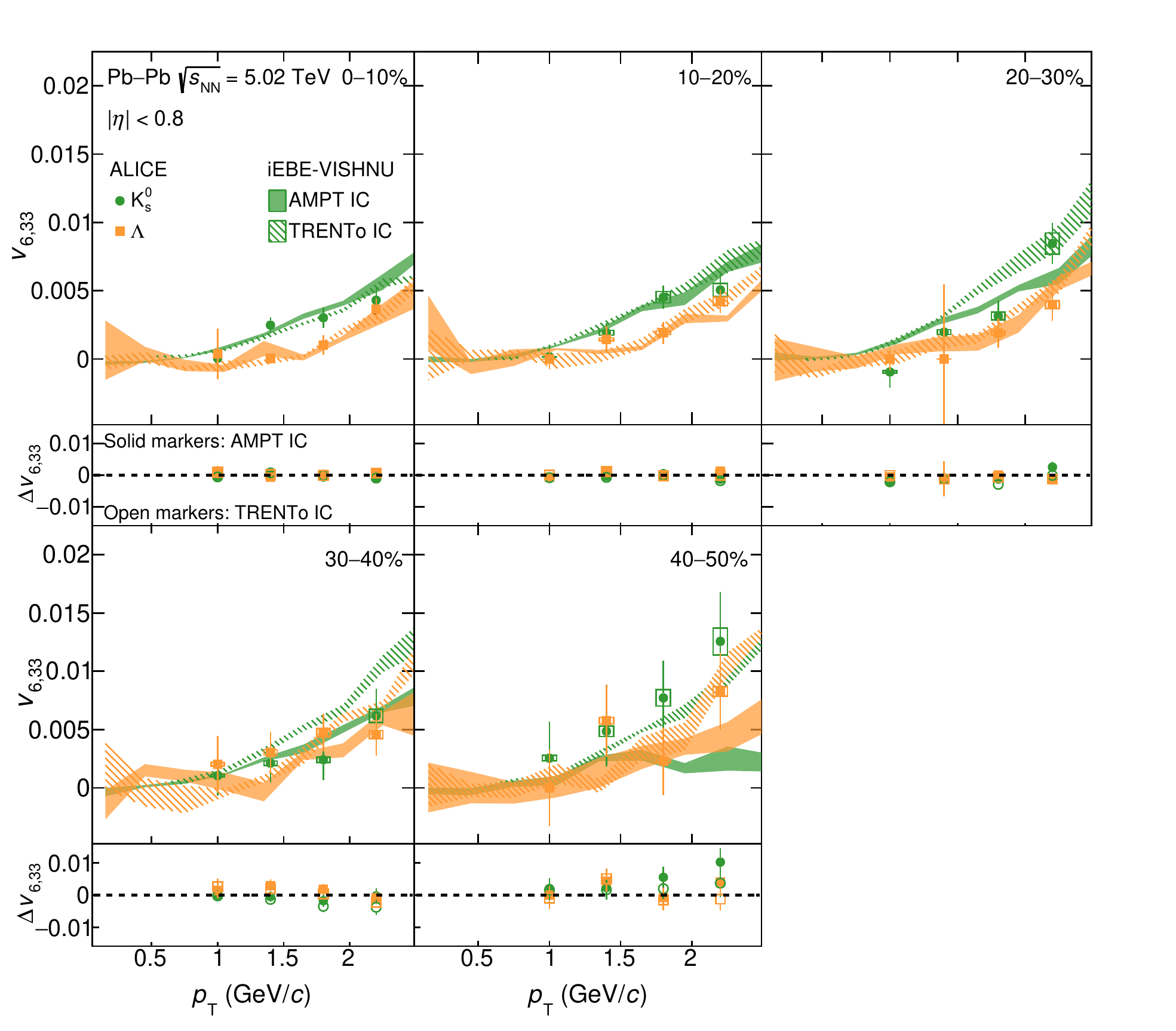}
\end{center}
\caption{The \pT-differential $v_{6,33}$ of \Ks~and \lambdas~in the 0--10\% up to 40--50\% centrality intervals of Pb--Pb collisions at \sNN compared with iEBE-VISHNU hybrid models with two different sets of initial parameters: AMPT initial conditions ($\eta/s$= 0.08 and $\zeta/s$ = 0) shown as solid bands and \trento~initial conditions ($\eta/s({\rm T})$ and $\zeta/s({\rm T})$) as hatched bands. The bottom panels show the difference between the measurements and each model. Statistical and systematic uncertainties are shown as bars and boxes, respectively.}
\label{v633_model_KL}
\end{figure}

To achieve additional constraints on the initial conditions and transport properties of the system and test the validity of these hydrodynamic models, a comparison is performed between the measured \pT-dependent non-linear flow modes for \pion, \kaon, \proton, \Ks~and \lambdas~with the same two hydrodynamical calculations reported in \cite{Zhao:2017yhj}. Figures \ref{v422_model}--\ref{v6222_model} present the comparison between the measurements and the two model predictions for the \pT-differential $v_{4,22}$, $v_{5,32}$, $v_{6,33}$ and $v_{6,222}$, respectively, for \pion, \kaon~and \proton~and Figs. \ref{v422_model_KL}--\ref{v633_model_KL} present these comparisons for the \pT-differential $v_{4,22}$, $v_{5,32}$ and $v_{6,33}$ for \Ks~and \lambdas~for the 0--10\% up to 50--60\% centrality interval (40--50\% centrality interval for $v_{6,33}$) of Pb--Pb collisions at \sNN. The solid bands show the AMPT model and the hatched bands represent the \trento~ calculations. The bottom panels in each plot in Figs. \ref{v422_model}--\ref{v633_model_KL} show the difference between the models and the measurement. Both \trento~and AMPT reproduce the mass ordering feature at $p_{\rm{T}}<2.5$ \GeV~for all non-linear flow modes. In particular, the comparison between the models and the measurements of $v_{4,22}$ reveals that \trento~reproduces the data very well from the 0--10\% up to 30--40\% centrality interval and fails to reproduce the measurements for the remaining more peripheral centrality intervals. On the other hand, AMPT overestimates the measurements from the 0--10\% up to 30--40\% centrality interval. For the 40--50\% centrality interval, it reproduces the measurements for all particle species except \pion, where it slightly underestimates the results. For more peripheral collisions, it reproduces the \kaon, \proton~and \lambdas~measurements and underestimates the results for \pion~and \Ks. 

In a similar attempt to the comparison between the $v_{n}$ measurement and the model calculation in Tab. \ref{ModelDataComparisonflow}, the performance of these models were further studied for $v_{n,mk}$ by taking the relative ratios between each model and the measurements of \pion, \kaon~and \proton. These relative ratios are summarised in Tab. \ref{ModelDataComparisonNLflow} where \trento~ calculations reproduce $v_{4,22}$ better than AMPT by $\sim$7\%. Comparisons between Tab. \ref{ModelDataComparisonNLflow} and \ref{ModelDataComparisonflow} show that the AMPT calculations reproduce $v_{4,22}$ with $\sim$20\% higher discrepancy on average compared to $v_{4}$, and, the \trento~calculations perform equally well for $v_{4,22}$ as for $v_{4}$. It is necessary to stress, however, that the non-linear flow modes have smaller magnitudes with respect to $v_{n}$ and any discrepancy between the models and the data becomes magnified in the ratios reported in Tab. \ref{ModelDataComparisonNLflow}. 

For $v_{5,32}$, the comparison is different, with the \trento~ predictions overestimating the measurements for all centrality intervals, and AMPT reproducing the data better than \trento. The AMPT model overestimates the measurements from the 0--10\% to 20--30\% centrality interval. It underestimates the measurements of \pion, \kaon~and \proton~for more peripheral collisions while it reproduces the measurements of \Ks~and \lambdas~relatively well up to the 40--50\% centrality interval. These comparisons are reflected in Tab. \ref{ModelDataComparisonNLflow} where AMPT performs on average 20--27\% better than \trento~for \pion, \kaon~and \proton. 

For $v_{6,33}$, both models reproduce the data for the 0--10\% centrality interval. For the 10--20\% up to 30--40\% centrality interval, AMPT reproduces the data while \trento~overestimates the measurements. Finally, the comparison with $v_{6,222}$ shows an agreement between both models and the measurements of \pion, \kaon~and \proton~at 0--10\% up to 30--40\% centrality intervals  \footnote{The ratios reported for $v_{6,33}$ and $v_{6,222}$ in Tab. \ref{ModelDataComparisonNLflow} are not to be taken at face value as the magnitudes of these two non-linear flow modes are almost zero.}.

All in all, this study shows larger differences between the model calculations and the $v_{n,mk}$ measurements with respect to that of $v_{\rm n}$, indicating a larger sensitivity to the initial conditions and transport properties for the non-linear flow modes. As a result, it is useful to tune the input parameters of hydrodynamic models considering also the non-linear flow measurements. 

\newpage
\newpage

\section{Summary}
\label{Sec:conclusion}

In this article, the measurements of the non-linear flow modes, $v_{4,22}$, $v_{5,32}$, $v_{6,222}$ and $v_{6,33}$ are for the first time reported as a function of transverse momentum for different particle species, i.e. \pion, \kaon, \Ks, \proton, \lambdas~and $\phi$-meson. The results are presented in a wide range of centrality intervals from 0--5\% up to 50--60\% in Pb--Pb collisions at \sNN. The magnitude of the non-linear flow modes, $v_{\rm n,mk}$, were obtained with a multi-particle correlation technique, namely the generic framework, selecting the identified hadron under study and the reference flow particles from different, non-overlapping pseudorapidity regions.  

The measured $v_{4,22}$, $v_{5,32}$ and $v_{6,222}$ exhibit a distinct centrality dependence. This centrality dependence originates from the contribution of initial state eccentricity, $\varepsilon_{2}$, as shown in Eq. \ref{Eq:V4V5V6}. As expected, $v_{6,33}$ does not exhibit a considerable centrality dependence since $\varepsilon_{3}$ quantifies primarily the event-by-event fluctuations of the initial energy density profile. This is supported by the relatively large magnitude of $v_{6,33}$ in the most-central collisions (0--5\%). A clear mass ordering is observed in the low \pT~region (\pT$< 2.5$ \GeV). A closer comparison between $v_{4}$ and $v_{4,22}$ shows that this mass ordering seems slightly larger for $v_{4,22}$ than $v_{4}$ at very low \pT~(\pT<0.8 \GeV). 
In the intermediate \pT~region (\pT$> 2.5$ \GeV), a particle type grouping is observed where the magnitude of the non-linear modes for baryons is larger than for mesons similar to observations in $v_{n}$ measurements. The NCQ scaling holds at an approximate level of $\pm 20$\% within the current level of statistical and systematic uncertainties, similar to that of the anisotropic flow coefficients \cite{Acharya:2018zuq}. 

The comparison of two models based on the iEBE-VISHNU hybrid model, with two different initial conditions (AMPT and \trento) and transport properties shows that neither of the models is able to fully describe the measurements. The quality of the model description depends on the centrality percentile and particle species similar to the model-data comparisons of the anisotropic flow coefficients \cite{Acharya:2018zuq}. The measurements are better predicted by the models in more central collisions. All in all, the model using AMPT initial conditions ($\eta/s = 0.08$ and $\zeta/s =0$) exhibits a magnitude and shape closer to the measurements. As a result, in order to further constrain the values of the transport properties and the initial conditions of the system, it is necessary to tune the input parameters of future hydrodynamic calculations attempting to describe these measurements.


\newpage
\newenvironment{acknowledgement}{\relax}{\relax}
\begin{acknowledgement}
\section*{Acknowledgements}

The ALICE Collaboration would like to thank all its engineers and technicians for their invaluable contributions to the construction of the experiment and the CERN accelerator teams for the outstanding performance of the LHC complex.
The ALICE Collaboration gratefully acknowledges the resources and support provided by all Grid centres and the Worldwide LHC Computing Grid (WLCG) collaboration.
The ALICE Collaboration acknowledges the following funding agencies for their support in building and running the ALICE detector:
A. I. Alikhanyan National Science Laboratory (Yerevan Physics Institute) Foundation (ANSL), State Committee of Science and World Federation of Scientists (WFS), Armenia;
Austrian Academy of Sciences, Austrian Science Fund (FWF): [M 2467-N36] and Nationalstiftung f\"{u}r Forschung, Technologie und Entwicklung, Austria;
Ministry of Communications and High Technologies, National Nuclear Research Center, Azerbaijan;
Conselho Nacional de Desenvolvimento Cient\'{\i}fico e Tecnol\'{o}gico (CNPq), Financiadora de Estudos e Projetos (Finep), Funda\c{c}\~{a}o de Amparo \`{a} Pesquisa do Estado de S\~{a}o Paulo (FAPESP) and Universidade Federal do Rio Grande do Sul (UFRGS), Brazil;
Ministry of Education of China (MOEC) , Ministry of Science \& Technology of China (MSTC) and National Natural Science Foundation of China (NSFC), China;
Ministry of Science and Education and Croatian Science Foundation, Croatia;
Centro de Aplicaciones Tecnol\'{o}gicas y Desarrollo Nuclear (CEADEN), Cubaenerg\'{\i}a, Cuba;
Ministry of Education, Youth and Sports of the Czech Republic, Czech Republic;
The Danish Council for Independent Research | Natural Sciences, the VILLUM FONDEN and Danish National Research Foundation (DNRF), Denmark;
Helsinki Institute of Physics (HIP), Finland;
Commissariat \`{a} l'Energie Atomique (CEA), Institut National de Physique Nucl\'{e}aire et de Physique des Particules (IN2P3) and Centre National de la Recherche Scientifique (CNRS) and R\'{e}gion des  Pays de la Loire, France;
Bundesministerium f\"{u}r Bildung und Forschung (BMBF) and GSI Helmholtzzentrum f\"{u}r Schwerionenforschung GmbH, Germany;
General Secretariat for Research and Technology, Ministry of Education, Research and Religions, Greece;
National Research, Development and Innovation Office, Hungary;
Department of Atomic Energy Government of India (DAE), Department of Science and Technology, Government of India (DST), University Grants Commission, Government of India (UGC) and Council of Scientific and Industrial Research (CSIR), India;
Indonesian Institute of Science, Indonesia;
Centro Fermi - Museo Storico della Fisica e Centro Studi e Ricerche Enrico Fermi and Istituto Nazionale di Fisica Nucleare (INFN), Italy;
Institute for Innovative Science and Technology , Nagasaki Institute of Applied Science (IIST), Japanese Ministry of Education, Culture, Sports, Science and Technology (MEXT) and Japan Society for the Promotion of Science (JSPS) KAKENHI, Japan;
Consejo Nacional de Ciencia (CONACYT) y Tecnolog\'{i}a, through Fondo de Cooperaci\'{o}n Internacional en Ciencia y Tecnolog\'{i}a (FONCICYT) and Direcci\'{o}n General de Asuntos del Personal Academico (DGAPA), Mexico;
Nederlandse Organisatie voor Wetenschappelijk Onderzoek (NWO), Netherlands;
The Research Council of Norway, Norway;
Commission on Science and Technology for Sustainable Development in the South (COMSATS), Pakistan;
Pontificia Universidad Cat\'{o}lica del Per\'{u}, Peru;
Ministry of Science and Higher Education and National Science Centre, Poland;
Korea Institute of Science and Technology Information and National Research Foundation of Korea (NRF), Republic of Korea;
Ministry of Education and Scientific Research, Institute of Atomic Physics and Ministry of Research and Innovation and Institute of Atomic Physics, Romania;
Joint Institute for Nuclear Research (JINR), Ministry of Education and Science of the Russian Federation, National Research Centre Kurchatov Institute, Russian Science Foundation and Russian Foundation for Basic Research, Russia;
Ministry of Education, Science, Research and Sport of the Slovak Republic, Slovakia;
National Research Foundation of South Africa, South Africa;
Swedish Research Council (VR) and Knut \& Alice Wallenberg Foundation (KAW), Sweden;
European Organization for Nuclear Research, Switzerland;
Suranaree University of Technology (SUT), National Science and Technology Development Agency (NSDTA) and Office of the Higher Education Commission under NRU project of Thailand, Thailand;
Turkish Atomic Energy Agency (TAEK), Turkey;
National Academy of  Sciences of Ukraine, Ukraine;
Science and Technology Facilities Council (STFC), United Kingdom;
National Science Foundation of the United States of America (NSF) and United States Department of Energy, Office of Nuclear Physics (DOE NP), United States of America.
\end{acknowledgement}

\newpage

\bibliographystyle{utphys}   
\bibliography{main}

\newpage
\appendix
\section{Additional figures}

\subsection{$\rm{KE_{T}}$ scaling}
\label{Subsubsection:KETscaling}

\begin{figure}[htb]
\begin{center}
\includegraphics[scale=0.82]{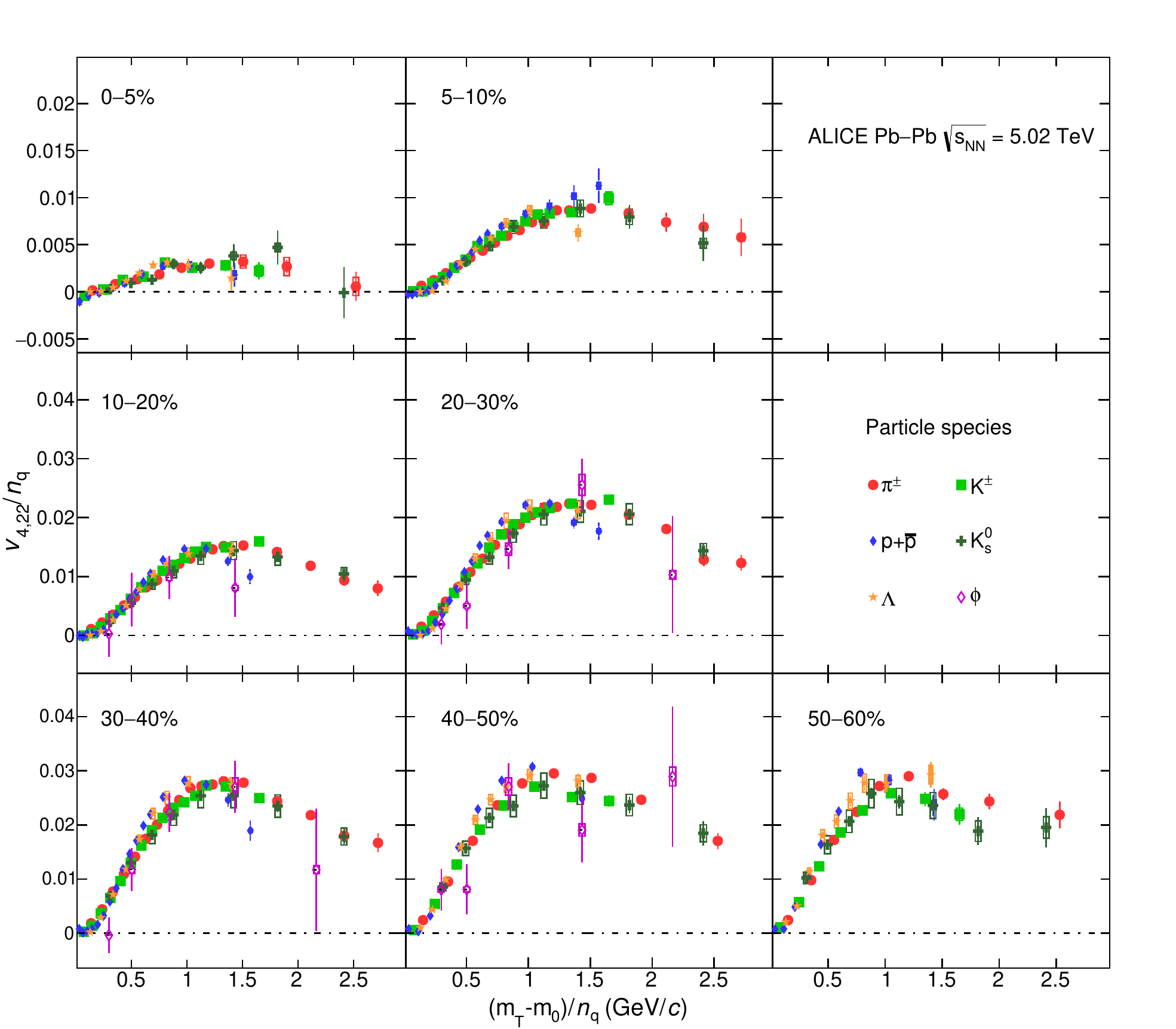}

\end{center}
\caption{The $(m_{\rm{T}} - m_{0})/n_{q}$-dependence of $v_{4,22}/n_{q}$ for different particle species grouped into different centrality intervals of Pb--Pb collisions \sNN. Statistical and systematic uncertainties are shown as bars and boxes, respectively. It is seen that the $\rm KE_{T}$ scaling holds for $v_{4,22}$ at an approximate level.}
\label{v422_KET}
\end{figure}

\begin{figure}[htb]
\begin{center}
\includegraphics[scale=0.82]{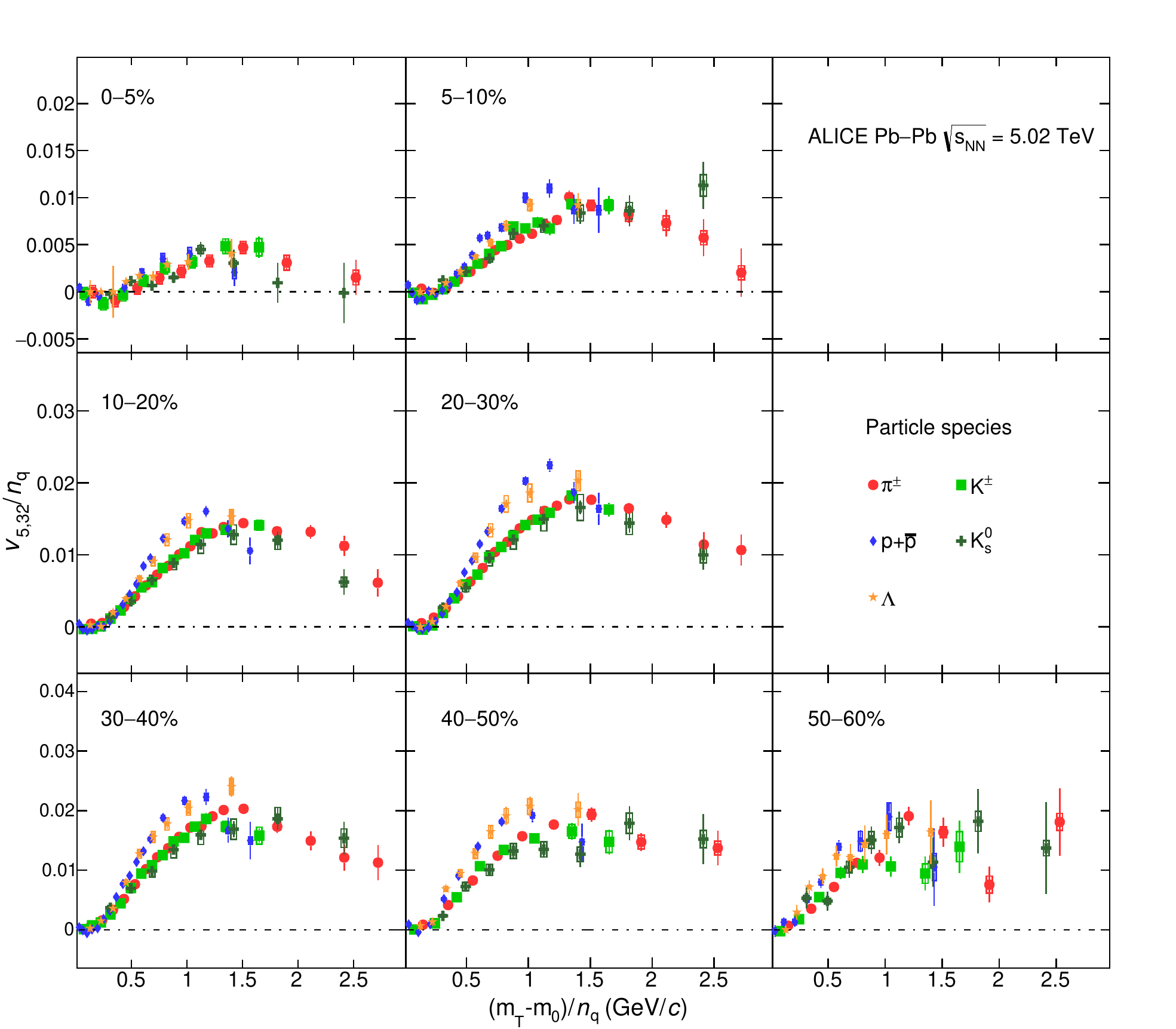}
\end{center}
\caption{The $(m_{\rm{T}} - m_{0})/n_{q}$-dependence of $v_{5,32}/n_{q}$ for different particle species grouped into different centrality intervals of Pb--Pb collisions \sNN. Statistical and systematic uncertainties are shown as bars and boxes, respectively. It is seen that the $\rm KE_{T}$ scaling holds for $v_{5,32}$ at an approximate level.}
\label{v523_KET}
\end{figure}

\begin{figure}[htb]
\begin{center}
\includegraphics[scale=0.82]{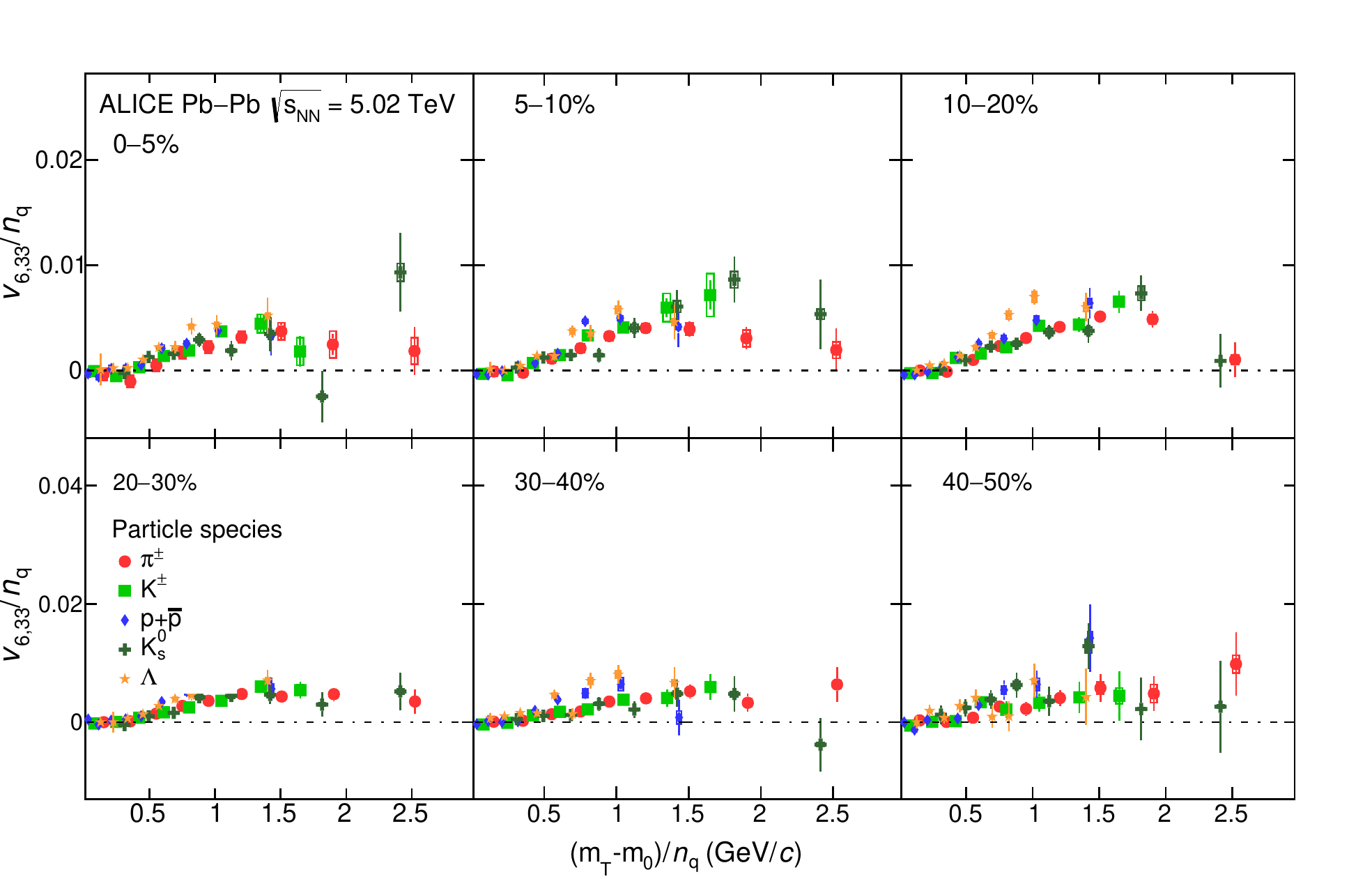}
\end{center}
\caption{The $(m_{\rm{T}} - m_{0})/n_{q}$-dependence of $v_{6,33}/n_{q}$ for different particle species grouped into different centrality intervals of Pb--Pb collisions \sNN. Statistical and systematic uncertainties are shown as bars and boxes, respectively. It is seen that the $\rm KE_{T}$ scaling holds for $v_{6,33}$ at an approximate level.}
\label{v633_KET}
\end{figure}

\begin{figure}[htb]
\begin{center}
\includegraphics[scale=0.82]{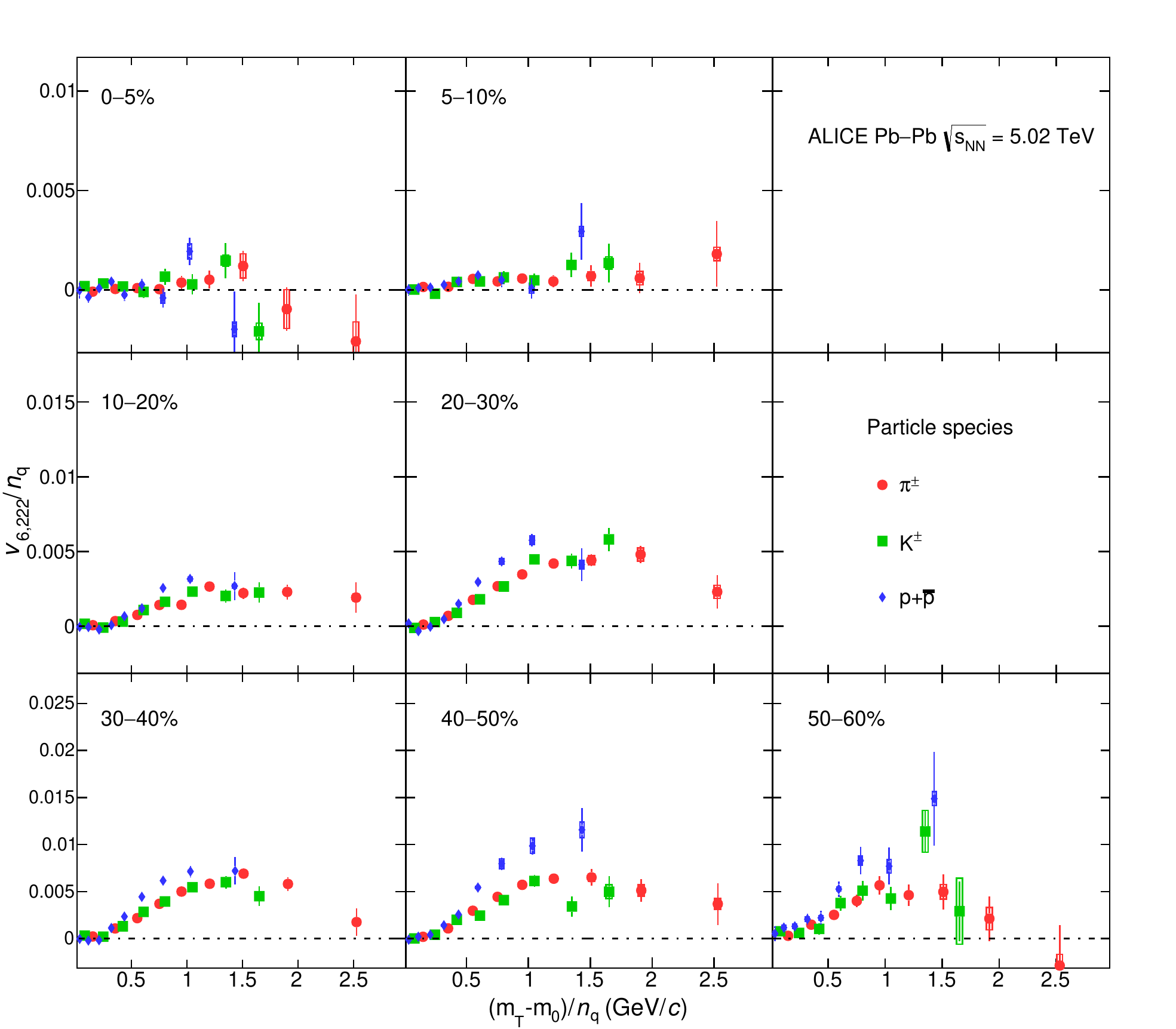}
\end{center}
\caption{The $(m_{\rm{T}} - m_{0})/n_{q}$-dependence of $v_{6,222}/n_{q}$ for different particle species grouped into different centrality intervals of Pb--Pb collisions \sNN. Statistical and systematic uncertainties are shown as bars and boxes, respectively. It is seen that the $\rm KE_{T}$ scaling holds for $v_{6,222}$ at an approximate level.}
\label{v6222_KET}
\end{figure}

One suggestion to further study the scaling properties of flow coefficients was to extend the scaling to lower \pT~values by studying the transverse kinetic energy dependence of anisotropic flow harmonics. Transverse kinetic energy is defined as ${\rm KE}_{\rm{T}} = m_{\rm{T}} - m_{\rm{0}}$, where $m_{\rm{T}}= \sqrt{m_{0}^2 + p_{\rm{T}}^2}$ is the transverse mass. Figures \ref{v422_KET}, \ref{v523_KET}, \ref{v633_KET} and \ref{v6222_KET} present ${\rm KE}_{\rm{T}}$ scaling for $v_{4,22}$, $v_{5,32}$, $v_{6,33}$ and $v_{6,222}$ respectively, for \pion, \kaon, \proton, \Ks, \lambdas~and $\phi$-meson grouped in different centrality intervals.

%
%

\section{The ALICE Collaboration}
\label{app:collab}

\begingroup
\small
\begin{flushleft}
S.~Acharya\Irefn{org141}\And 
D.~Adamov\'{a}\Irefn{org94}\And 
A.~Adler\Irefn{org74}\And 
J.~Adolfsson\Irefn{org80}\And 
M.M.~Aggarwal\Irefn{org99}\And 
G.~Aglieri Rinella\Irefn{org33}\And 
M.~Agnello\Irefn{org30}\And 
N.~Agrawal\Irefn{org10}\textsuperscript{,}\Irefn{org53}\And 
Z.~Ahammed\Irefn{org141}\And 
S.~Ahmad\Irefn{org16}
% ...
\And
\renewcommand\labelenumi{\textsuperscript{\theenumi}~}

\section*{Affiliation notes}
\renewcommand\theenumi{\roman{enumi}}
\begin{Authlist}
\item \Adef{org*}Deceased
\item \Adef{orgI}Dipartimento DET del Politecnico di Torino, Turin, Italy
\item \Adef{orgII}M.V. Lomonosov Moscow State University, D.V. Skobeltsyn Institute of Nuclear, Physics, Moscow, Russia
\item \Adef{orgIII}Department of Applied Physics, Aligarh Muslim University, Aligarh, India
\item \Adef{orgIV}Institute of Theoretical Physics, University of Wroclaw, Poland
\end{Authlist}

\section*{Collaboration Institutes}
\renewcommand\theenumi{\arabic{enumi}~}
\begin{Authlist}
\item \Idef{org1}A.I. Alikhanyan National Science Laboratory (Yerevan Physics Institute) Foundation, Yerevan, Armenia
\item \Idef{org2}Bogolyubov Institute for Theoretical Physics, National Academy of Sciences of Ukraine, Kiev, Ukraine
\item \Idef{org3}Bose Institute, Department of Physics  and Centre for Astroparticle Physics and Space Science (CAPSS), Kolkata, India
\item \Idef{org4}Budker Institute for Nuclear Physics, Novosibirsk, Russia
\item \Idef{org5}California Polytechnic State University, San Luis Obispo, California, United States
\item \Idef{org6}Central China Normal University, Wuhan, China
\item \Idef{org7}Centre de Calcul de l'IN2P3, Villeurbanne, Lyon, France
\item \Idef{org8}Centro de Aplicaciones Tecnol\'{o}gicas y Desarrollo Nuclear (CEADEN), Havana, Cuba
\item \Idef{org9}Centro de Investigaci\'{o}n y de Estudios Avanzados (CINVESTAV), Mexico City and M\'{e}rida, Mexico
\item \Idef{org10}Centro Fermi - Museo Storico della Fisica e Centro Studi e Ricerche ``Enrico Fermi', Rome, Italy
\item \Idef{org11}Chicago State University, Chicago, Illinois, United States
\item \Idef{org12}China Institute of Atomic Energy, Beijing, China
\item \Idef{org13}Comenius University Bratislava, Faculty of Mathematics, Physics and Informatics, Bratislava, Slovakia
\item \Idef{org14}COMSATS University Islamabad, Islamabad, Pakistan
\item \Idef{org15}Creighton University, Omaha, Nebraska, United States
\item \Idef{org16}Department of Physics, Aligarh Muslim University, Aligarh, India
\item \Idef{org17}Department of Physics, Pusan National University, Pusan, Republic of Korea
\item \Idef{org18}Department of Physics, Sejong University, Seoul, Republic of Korea
\item \Idef{org19}Department of Physics, University of California, Berkeley, California, United States
\item \Idef{org20}Department of Physics, University of Oslo, Oslo, Norway
\item \Idef{org21}Department of Physics and Technology, University of Bergen, Bergen, Norway
\item \Idef{org22}Dipartimento di Fisica dell'Universit\`{a} 'La Sapienza' and Sezione INFN, Rome, Italy
\item \Idef{org23}Dipartimento di Fisica dell'Universit\`{a} and Sezione INFN, Cagliari, Italy
\item \Idef{org24}Dipartimento di Fisica dell'Universit\`{a} and Sezione INFN, Trieste, Italy
\item \Idef{org25}Dipartimento di Fisica dell'Universit\`{a} and Sezione INFN, Turin, Italy
\item \Idef{org26}Dipartimento di Fisica e Astronomia dell'Universit\`{a} and Sezione INFN, Bologna, Italy
\item \Idef{org27}Dipartimento di Fisica e Astronomia dell'Universit\`{a} and Sezione INFN, Catania, Italy
\item \Idef{org28}Dipartimento di Fisica e Astronomia dell'Universit\`{a} and Sezione INFN, Padova, Italy
\item \Idef{org29}Dipartimento di Fisica `E.R.~Caianiello' dell'Universit\`{a} and Gruppo Collegato INFN, Salerno, Italy
\item \Idef{org30}Dipartimento DISAT del Politecnico and Sezione INFN, Turin, Italy
\item \Idef{org31}Dipartimento di Scienze e Innovazione Tecnologica dell'Universit\`{a} del Piemonte Orientale and INFN Sezione di Torino, Alessandria, Italy
\item \Idef{org32}Dipartimento Interateneo di Fisica `M.~Merlin' and Sezione INFN, Bari, Italy
\item \Idef{org33}European Organization for Nuclear Research (CERN), Geneva, Switzerland
\item \Idef{org34}Faculty of Electrical Engineering, Mechanical Engineering and Naval Architecture, University of Split, Split, Croatia
\item \Idef{org35}Faculty of Engineering and Science, Western Norway University of Applied Sciences, Bergen, Norway
\item \Idef{org36}Faculty of Nuclear Sciences and Physical Engineering, Czech Technical University in Prague, Prague, Czech Republic
\item \Idef{org37}Faculty of Science, P.J.~\v{S}af\'{a}rik University, Ko\v{s}ice, Slovakia
\item \Idef{org38}Frankfurt Institute for Advanced Studies, Johann Wolfgang Goethe-Universit\"{a}t Frankfurt, Frankfurt, Germany
\item \Idef{org39}Fudan University, Shanghai, China
\item \Idef{org40}Gangneung-Wonju National University, Gangneung, Republic of Korea
\item \Idef{org41}Gauhati University, Department of Physics, Guwahati, India
\item \Idef{org42}Helmholtz-Institut f\"{u}r Strahlen- und Kernphysik, Rheinische Friedrich-Wilhelms-Universit\"{a}t Bonn, Bonn, Germany
\item \Idef{org43}Helsinki Institute of Physics (HIP), Helsinki, Finland
\item \Idef{org44}High Energy Physics Group,  Universidad Aut\'{o}noma de Puebla, Puebla, Mexico
\item \Idef{org45}Hiroshima University, Hiroshima, Japan
\item \Idef{org46}Hochschule Worms, Zentrum  f\"{u}r Technologietransfer und Telekommunikation (ZTT), Worms, Germany
\item \Idef{org47}Horia Hulubei National Institute of Physics and Nuclear Engineering, Bucharest, Romania
\item \Idef{org48}Indian Institute of Technology Bombay (IIT), Mumbai, India
\item \Idef{org49}Indian Institute of Technology Indore, Indore, India
\item \Idef{org50}Indonesian Institute of Sciences, Jakarta, Indonesia
\item \Idef{org51}INFN, Laboratori Nazionali di Frascati, Frascati, Italy
\item \Idef{org52}INFN, Sezione di Bari, Bari, Italy
\item \Idef{org53}INFN, Sezione di Bologna, Bologna, Italy
\item \Idef{org54}INFN, Sezione di Cagliari, Cagliari, Italy
\item \Idef{org55}INFN, Sezione di Catania, Catania, Italy
\item \Idef{org56}INFN, Sezione di Padova, Padova, Italy
\item \Idef{org57}INFN, Sezione di Roma, Rome, Italy
\item \Idef{org58}INFN, Sezione di Torino, Turin, Italy
\item \Idef{org59}INFN, Sezione di Trieste, Trieste, Italy
\item \Idef{org60}Inha University, Incheon, Republic of Korea
\item \Idef{org61}Institut de Physique Nucl\'{e}aire d'Orsay (IPNO), Institut National de Physique Nucl\'{e}aire et de Physique des Particules (IN2P3/CNRS), Universit\'{e} de Paris-Sud, Universit\'{e} Paris-Saclay, Orsay, France
\item \Idef{org62}Institute for Nuclear Research, Academy of Sciences, Moscow, Russia
\item \Idef{org63}Institute for Subatomic Physics, Utrecht University/Nikhef, Utrecht, Netherlands
\item \Idef{org64}Institute of Experimental Physics, Slovak Academy of Sciences, Ko\v{s}ice, Slovakia
\item \Idef{org65}Institute of Physics, Homi Bhabha National Institute, Bhubaneswar, India
\item \Idef{org66}Institute of Physics of the Czech Academy of Sciences, Prague, Czech Republic
\item \Idef{org67}Institute of Space Science (ISS), Bucharest, Romania
\item \Idef{org68}Institut f\"{u}r Kernphysik, Johann Wolfgang Goethe-Universit\"{a}t Frankfurt, Frankfurt, Germany
\item \Idef{org69}Instituto de Ciencias Nucleares, Universidad Nacional Aut\'{o}noma de M\'{e}xico, Mexico City, Mexico
\item \Idef{org70}Instituto de F\'{i}sica, Universidade Federal do Rio Grande do Sul (UFRGS), Porto Alegre, Brazil
\item \Idef{org71}Instituto de F\'{\i}sica, Universidad Nacional Aut\'{o}noma de M\'{e}xico, Mexico City, Mexico
\item \Idef{org72}iThemba LABS, National Research Foundation, Somerset West, South Africa
\item \Idef{org73}Jeonbuk National University, Jeonju, Republic of Korea
\item \Idef{org74}Johann-Wolfgang-Goethe Universit\"{a}t Frankfurt Institut f\"{u}r Informatik, Fachbereich Informatik und Mathematik, Frankfurt, Germany
\item \Idef{org75}Joint Institute for Nuclear Research (JINR), Dubna, Russia
\item \Idef{org76}Korea Institute of Science and Technology Information, Daejeon, Republic of Korea
\item \Idef{org77}KTO Karatay University, Konya, Turkey
\item \Idef{org78}Laboratoire de Physique Subatomique et de Cosmologie, Universit\'{e} Grenoble-Alpes, CNRS-IN2P3, Grenoble, France
\item \Idef{org79}Lawrence Berkeley National Laboratory, Berkeley, California, United States
\item \Idef{org80}Lund University Department of Physics, Division of Particle Physics, Lund, Sweden
\item \Idef{org81}Nagasaki Institute of Applied Science, Nagasaki, Japan
\item \Idef{org82}Nara Women{'}s University (NWU), Nara, Japan
\item \Idef{org83}National and Kapodistrian University of Athens, School of Science, Department of Physics , Athens, Greece
\item \Idef{org84}National Centre for Nuclear Research, Warsaw, Poland
\item \Idef{org85}National Institute of Science Education and Research, Homi Bhabha National Institute, Jatni, India
\item \Idef{org86}National Nuclear Research Center, Baku, Azerbaijan
\item \Idef{org87}National Research Centre Kurchatov Institute, Moscow, Russia
\item \Idef{org88}Niels Bohr Institute, University of Copenhagen, Copenhagen, Denmark
\item \Idef{org89}Nikhef, National institute for subatomic physics, Amsterdam, Netherlands
\item \Idef{org90}NRC Kurchatov Institute IHEP, Protvino, Russia
\item \Idef{org91}NRC «Kurchatov Institute»  - ITEP, Moscow, Russia
\item \Idef{org92}NRNU Moscow Engineering Physics Institute, Moscow, Russia
\item \Idef{org93}Nuclear Physics Group, STFC Daresbury Laboratory, Daresbury, United Kingdom
\item \Idef{org94}Nuclear Physics Institute of the Czech Academy of Sciences, \v{R}e\v{z} u Prahy, Czech Republic
\item \Idef{org95}Oak Ridge National Laboratory, Oak Ridge, Tennessee, United States
\item \Idef{org96}Ohio State University, Columbus, Ohio, United States
\item \Idef{org97}Petersburg Nuclear Physics Institute, Gatchina, Russia
\item \Idef{org98}Physics department, Faculty of science, University of Zagreb, Zagreb, Croatia
\item \Idef{org99}Physics Department, Panjab University, Chandigarh, India
\item \Idef{org100}Physics Department, University of Jammu, Jammu, India
\item \Idef{org101}Physics Department, University of Rajasthan, Jaipur, India
\item \Idef{org102}Physikalisches Institut, Eberhard-Karls-Universit\"{a}t T\"{u}bingen, T\"{u}bingen, Germany
\item \Idef{org103}Physikalisches Institut, Ruprecht-Karls-Universit\"{a}t Heidelberg, Heidelberg, Germany
\item \Idef{org104}Physik Department, Technische Universit\"{a}t M\"{u}nchen, Munich, Germany
\item \Idef{org105}Politecnico di Bari, Bari, Italy
\item \Idef{org106}Research Division and ExtreMe Matter Institute EMMI, GSI Helmholtzzentrum f\"ur Schwerionenforschung GmbH, Darmstadt, Germany
\item \Idef{org107}Rudjer Bo\v{s}kovi\'{c} Institute, Zagreb, Croatia
\item \Idef{org108}Russian Federal Nuclear Center (VNIIEF), Sarov, Russia
\item \Idef{org109}Saha Institute of Nuclear Physics, Homi Bhabha National Institute, Kolkata, India
\item \Idef{org110}School of Physics and Astronomy, University of Birmingham, Birmingham, United Kingdom
\item \Idef{org111}Secci\'{o}n F\'{\i}sica, Departamento de Ciencias, Pontificia Universidad Cat\'{o}lica del Per\'{u}, Lima, Peru
\item \Idef{org112}St. Petersburg State University, St. Petersburg, Russia
\item \Idef{org113}Stefan Meyer Institut f\"{u}r Subatomare Physik (SMI), Vienna, Austria
\item \Idef{org114}SUBATECH, IMT Atlantique, Universit\'{e} de Nantes, CNRS-IN2P3, Nantes, France
\item \Idef{org115}Suranaree University of Technology, Nakhon Ratchasima, Thailand
\item \Idef{org116}Technical University of Ko\v{s}ice, Ko\v{s}ice, Slovakia
\item \Idef{org117}Technische Universit\"{a}t M\"{u}nchen, Excellence Cluster 'Universe', Munich, Germany
\item \Idef{org118}The Henryk Niewodniczanski Institute of Nuclear Physics, Polish Academy of Sciences, Cracow, Poland
\item \Idef{org119}The University of Texas at Austin, Austin, Texas, United States
\item \Idef{org120}Universidad Aut\'{o}noma de Sinaloa, Culiac\'{a}n, Mexico
\item \Idef{org121}Universidade de S\~{a}o Paulo (USP), S\~{a}o Paulo, Brazil
\item \Idef{org122}Universidade Estadual de Campinas (UNICAMP), Campinas, Brazil
\item \Idef{org123}Universidade Federal do ABC, Santo Andre, Brazil
\item \Idef{org124}University of Cape Town, Cape Town, South Africa
\item \Idef{org125}University of Houston, Houston, Texas, United States
\item \Idef{org126}University of Jyv\"{a}skyl\"{a}, Jyv\"{a}skyl\"{a}, Finland
\item \Idef{org127}University of Liverpool, Liverpool, United Kingdom
\item \Idef{org128}University of Science and Technology of China, Hefei, China
\item \Idef{org129}University of South-Eastern Norway, Tonsberg, Norway
\item \Idef{org130}University of Tennessee, Knoxville, Tennessee, United States
\item \Idef{org131}University of the Witwatersrand, Johannesburg, South Africa
\item \Idef{org132}University of Tokyo, Tokyo, Japan
\item \Idef{org133}University of Tsukuba, Tsukuba, Japan
\item \Idef{org134}Universit\'{e} Clermont Auvergne, CNRS/IN2P3, LPC, Clermont-Ferrand, France
\item \Idef{org135}Universit\'{e} de Lyon, Universit\'{e} Lyon 1, CNRS/IN2P3, IPN-Lyon, Villeurbanne, Lyon, France
\item \Idef{org136}Universit\'{e} de Strasbourg, CNRS, IPHC UMR 7178, F-67000 Strasbourg, France, Strasbourg, France
\item \Idef{org137}Universit\'{e} Paris-Saclay Centre d'Etudes de Saclay (CEA), IRFU, D\'{e}partment de Physique Nucl\'{e}aire (DPhN), Saclay, France
\item \Idef{org138}Universit\`{a} degli Studi di Foggia, Foggia, Italy
\item \Idef{org139}Universit\`{a} degli Studi di Pavia, Pavia, Italy
\item \Idef{org140}Universit\`{a} di Brescia, Brescia, Italy
\item \Idef{org141}Variable Energy Cyclotron Centre, Homi Bhabha National Institute, Kolkata, India
\item \Idef{org142}Warsaw University of Technology, Warsaw, Poland
\item \Idef{org143}Wayne State University, Detroit, Michigan, United States
\item \Idef{org144}Westf\"{a}lische Wilhelms-Universit\"{a}t M\"{u}nster, Institut f\"{u}r Kernphysik, M\"{u}nster, Germany
\item \Idef{org145}Wigner Research Centre for Physics, Budapest, Hungary
\item \Idef{org146}Yale University, New Haven, Connecticut, United States
\item \Idef{org147}Yonsei University, Seoul, Republic of Korea
\end{Authlist}
\endgroup
  
\end{document}